\documentclass[11pt,a4paper]{article}

\usepackage{jheppub}
\usepackage{comment}

\usepackage[latin1]{inputenc} 
\usepackage[T1]{fontenc} 
\usepackage{graphicx} 
\usepackage{amsmath}
\usepackage{amsfonts}
\usepackage{amssymb}
\usepackage{latexsym}
\usepackage{feynmp}
\newcommand{\bea}{\begin{eqnarray}}
\newcommand{\eea}{\end{eqnarray}}

\title{Higgs sector in NMSSM with right-handed neutrinos and spontaneous R-parity violation}

\author{Katri Huitu}
\author{and Harri Waltari}

\affiliation{Department of Physics, and Helsinki Institute of Physics, P.O.Box 64, FIN-00014 University of Helsinki, Finland}

\emailAdd{katri.huitu@helsinki.fi}
\emailAdd{harri.waltari@helsinki.fi}

\abstract{
R-parity violation modifies the phenomenology of supersymmetric models considerably. We study a version of NMSSM,
which contains right-handed neutrinos and in which spontaneous R-parity violation is possible.
We study the ensuing effects of spontaneous breaking to the Higgs decay modes, taking into account the measured mass of the Higgs boson and experimental constraints, including rare decays.
We find that a possible light scalar, dominantly a sneutrino, helps to increase the Standard Model (SM)
Higgs-like scalar mass to the measured value.
At the same time, a lighter stop than in the MSSM is allowed.
The Higgs decay rates in the studied model can somewhat differ from the SM expectations, although the
most prominent difference is a universal 
suppression in the couplings due to the mixing of doublet scalars with singlets.
The charged, pseudoscalar, and other than the two lightest scalar Higgses are typically heavier than 1 TeV in
the parameter region where R-parity is spontaneously broken.
}

\arxivnumber{1405.5330}

\keywords{Supersymmetry phenomenology, Higgs boson, spontaneous R-parity violation}

\begin{document}\setlength{\unitlength}{1mm}

\preprint{HIP-2014-09/TH}

\maketitle

\section{Introduction}

Supersymmetry (SUSY) remains as the best motivated extension of the Standard Model (SM). 
However, so far all direct searches for superpartners have failed and thus there is no direct evidence for SUSY. 
The big discovery in the Large Hadron Collider (LHC) experiments, at CMS \cite{Chatrchyan:2012ufa} and at ATLAS \cite{Aad:2012tfa}, is a Higgs boson, which has mass compatible with the 
expectations for the lightest CP-even Higgs in supersymmetric models.
Most interestingly, the Higgs branching ratios may help to
understand physics beyond the Standard Model.
The statistics is not enough to make definite conclusions yet, but the branching ratios do not show
big deviations from the SM.
The only branching ratio, which has been measured to be clearly different from the Standard
Model, is the Higgs decay to two photons \cite{ATLAS:2012ad,Chatrchyan:2012tw}, although 
the latest results both from ATLAS \cite{Aad:2014eha} and from CMS \cite{Khachatryan:2014ira} are consistent with the Standard Model at one sigma level. 
It is well known that in the minimal supersymmetric standard model (MSSM) it is necessary
to have either a heavy stop or large mixing between the stops to achieve the measured
mass for the Higgs boson.
This raises questions about fine tuning in the MSSM.

A long lasting problem in the SM and in the MSSM has been the 
neutrino mass generation.
In MSSM neutrinos remain massless similar to the SM, so extensions to the MSSM are necessary. 
One possible way to explain the nonvanishing neutrino masses is to discard the assumption of 
exact conservation of the R-parity \cite{Salam:1974xa,Fayet:1974pd,Farrar:1978xj,Hall:1983id}, $R_P=(-1)^{3B-L+2s}$, where $B$ is baryon
number, $L$ is lepton number, and $s$ is spin of the particle.
Phenomenological implications of R-parity violation have been extensively studied, see 
\cite{Barbier:2004ez}.
If R-parity is not required, the MSSM superpotential breaks both baryon and lepton number conservation 
and leads to a much too fast proton decay unless couplings are tiny. 
However, there are models in which the R-parity breaking can naturally happen via breaking
only the lepton number and thus the proton life time is not affected.
This is the case when R-parity is spontaneously violated \cite{Aulakh:1982yn,Hayashi:1984rd,Mohapatra:1986aw, Masiero:1990uj}.
Also in this case the phenomenology and constraints on parameters have been studied, 
see {\it e.g.} \cite{Frank:2001tr}.
The R-parity violation (RPV) allows terms which can generate neutrino masses \cite{Grossman:1998py,Hirsch:2000ef,Davidson:2000uc,Abada:2002ju}.
Another much studied method to generate small neutrino masses is
by the seesaw mechanism \cite{Mohapatra:1979ia,Valle}
when right-handed neutrinos are included in the model.
The next-to-minimal supersymmetric standard model, NMSSM, is a simple extension of
the MSSM, including an extra scalar compared to the MSSM and with a $\mathbb{Z}_3$ symmetry
imposed.
An example of an NMSSM-type model with right-handed neutrinos where the seesaw mechanism 
with spontaneous R-parity breaking can produce 
the measured neutrino mass differences and mixing angles is given in 
\cite{Kitano:1999qb, Frank:2005tn, Frank:2007un}.
In this work we will use an NMSSM-type model with right-handed neutrinos 
as our framework.

When R-parity is broken, the commonly studied candidates for cold dark matter,
namely the neutralino and the right-handed sneutrino, see  \cite{Cao:2011re,Vasquez:2012hn,Cerdeno:2009dv,Cerdeno:2011qv,Huitu:2012rd}
for NMSSM,
are not stable and thus neither can be the SUSY candidate for cold dark matter because of the couplings to the SM particles. 
If the gravitino were the lightest supersymmetric particle (LSP), it could have a lifetime longer than the age of the Universe due to the Planck scale suppressed couplings, see \cite{Borgani:1996ag,Takayama:2000uz,JeanLouis:2009du,Luo:2010he}.
Gravitino is the LSP in gauge mediated supersymmetry breaking models, but their characteristic feature is small $A$-terms. 
As is shown later,
large $A$-terms are necessary both to lift the lightest Higgs boson mass to the measured value and to break R-parity spontaneously in the model considered in our work. 
However, recently it was proposed that in a nonminimal model for gauge mediation large $A$-terms can be generated \cite{Craig:2012xp}. 
One may also assume that a completely unknown sector is responsible for the dark matter.

The experimental limits for sparticle masses differ for R-parity conserving and violating cases,
since many of the methods of the R-parity conserving SUSY searches do not apply for RPV case.
In particular, the missing energy is significantly softened compared to the R-parity conserving model,
since the LSP decays either through couplings to the SM particles or through mixing with the SM particles,
and missing energy is not anymore one of the important characteristics of the model, see \textit{e.g.} \cite{Evans:2012bf,Graham:2014vya,CMS:2013qda,ATLAS:2012dp}. 
When R-parity is violated, both electric and color charges are possible for the LSP.

In this work we concentrate on the Higgs sector in the case of spontaneous R-parity breaking.
We study the neutral scalar particles and their decays in a model with spontaneous R-parity violation, 
where R-parity violation is generated by a VEV of one or several sneutrino f\mbox{}ields. 
The spontaneous breaking affects a number of couplings in the model.
Thus, contrary to the explicit breaking, it is not possible to choose one of the couplings to be the dominant one, but 
the couplings are determined by the sneutrino VEV and other parameters together.
This makes it necessary to check strict constraints from rare decays.
Similarly the constraints from experimental mass limits need to be satisfied.

We first review the relevant features of the model in Section 2, including the scalar sector of the model 
in general. The minimization conditions are found, and we discuss the scalar masses,
as well as consider constraints coming from fermion masses.
In Section 3 we scan over the relevant parameter space of the model, discuss constraints from rare decays
and consider the possible decay modes for the scalar with 125 GeV mass in the model.
This may be the lightest scalar, or there may be one or two lighter scalars, dominantly singlets.
We compute the 125 GeV Higgs production and subsequent decay to two photons.
We find that it is possible to have larger branching ratios than in the Standard Model, especially when the top coupling is enhanced, or bottom coupling is reduced from the Standard Model value. 
This, however, is not the typical situation in our scanned points. 
Even in the cases when the branching ratio is large compared to the SM, the rate is only moderately affected because of the simultaneous suppressed production.
In Section 4 we conclude.

\section{The model}
There are several models, which allow vacua with broken R-parity
\cite{Masiero:1990uj,Joshipura:1994wm,romao1,romao2,Chaichian:1991zt,Chaichian:1992ra,
Kuchimanchi:1993jg,Huitu:1994zm,Barger:2008wn}.
All viable ones introduce lepton number violation. If a global symmetry as lepton number is violated, the particle spectrum will have a Goldstone boson, called Majoron \cite{Chikashige:1980ui,Gelmini:1980re}. 
In this work we add explicitly lepton number violating terms to the superpotential which make the pseudo-Majoron massive.

We consider a model which has a superpotential \cite{Kitano:1999qb, Frank:2005tn, Frank:2007un}
\begin{multline}\label{eq:Kitano}
W=\sum_{i,j}\left( h^{d}_{ij}(\hat H_{d}\cdot\hat Q_{i})\hat D_{j} + h^{e}_{ij}(\hat H_{d}\cdot\hat L_{i})\hat E_{j}
+h^{u}_{ij}(\hat Q_{i}\cdot\hat H_{u})\hat U_{j}\right)\\
+\sum_{i,j}h^{\nu}_{ij}(\hat L_{i}\cdot\hat H_{u})\hat N_{j}+\lambda_{H}(\hat H_{d}\cdot \hat H_{u})
\hat \Phi+\sum_{i}\frac{\lambda_{N_{i}}}{2}\hat N_{i}^{2}\hat \Phi+\frac{\lambda_{\Phi}}{6}\hat \Phi^{3}.
\end{multline}
The first three terms are the MSSM superpotential without the so-called $\mu$-term. 
The $\hat N_i$ f\mbox{}ields are the neutrino singlet superf\mbox{}ields 
($L$=$-1$) and $\hat \Phi$ is a scalar singlet ($L$=0) superf\mbox{}ield.
The VEV of $\Phi$ produces the $\mu$-term of the superpotential as usual in the NMSSM, and the Majorana mass term for the right-handed neutrinos.

This superpotential does not introduce anything but a minimum number of f\mbox{}ields necessary to break R-parity spontaneously\footnote{If explicit R-parity violation is allowed, one may have sneutrino VEVs without the $L=0$ singlet also, see \cite{LopezFogliani:2005yw}.} and to have only trilinear couplings.
From the fourth term in the superpotential, it is obvious that when the right-handed sneutrino develops a VEV there will be a term that mimics the explicitly R-parity violating bilinear term proportional to $LH$.
The $\hat{N}^{2}\hat{\Phi}$-term breaks explicitly the lepton number, but not R-parity. 
In this model the pseudo-Majoron will be mostly singlet-like, so even if it were relatively light, it would not have been seen in Z boson decays.

Following the Ref.~\cite{Kitano:1999qb} we assume that the possible domain wall problems will be removed by 
nonrenormalizable terms or inflation.
 A $\mathbb{Z}_{3}$ symmetry has been imposed to the superpotential so that all couplings are dimensionless. 
When $\mathbb{Z}_{3}$ is spontaneously broken, a potential problem with cosmological domain walls appears \cite{Abel:1995wk,Maniatis:2009re}. 
Solutions to the problem have been proposed. 
One possibility is that 
the symmetry is explicitly broken by non-renormalizable terms so that a preferred vacuum exists. 
In general this will create huge tadpole terms to the singlet f\mbox{}ields so that the generation of a $\mu$-term with correct size requires fine-tuning.
It was shown in \cite{Abel:1996cr, Panagiotakopoulos:1998yw,Panagiotakopoulos:1999ah} that assuming a new discrete symmetry, which holds also for the non-renormalizable terms, only a tadpole term with a size of the supersymmetry breaking scale is generated for the singlet scalar.
The single resulting parameter in the softly supersymmetry breaking Lagrangian is denoted by $\xi$.
Thus, there are the following soft supersymmetry breaking terms in the Lagrangian
\begin{multline}\label{eq:Kitanosoft}
-\mathcal{L}_{\mathrm{soft}}=-\mathcal{L}_{\mathrm{soft}}^{\mathrm{MSSM'}}+\sum_{i}m_{N_i}^{2}|\tilde{N}_{i}|^{2}+m_{\Phi}^{2}|\Phi|^{2}+\left[ \sum_{i,j} A_{\nu}^{ij}(\tilde{L}_{i}\cdot H_{u})\tilde{N}_{j}\right.\\
\left. +A_{H}(H_{d}\cdot H_{u})\Phi +\sum_{i} \frac{1}{2}A_{N}\tilde{N}_{i}^{2}\Phi+\frac{1}{6}A_{\Phi}\Phi^{3}+\xi^{3}\Phi+\mathrm{h.c.}\right],
\end{multline}
where $\mathcal{L}_{\mathrm{soft}}^{\mathrm{MSSM'}}$ contains the MSSM soft terms without the $H_u$ and $H_d$ mixing bilinear term.

The model introduces new f\mbox{}ields and couplings (a total of 13 new complex parameters in the superpotential compared to the MSSM) but compared to explicit R-parity violation (48 new complex parameters in the superpotential) the model is economical.

\subsection{Neutral scalar potential}
Let's first study the situation with only one generation of SM fermions and singlet neutrinos. In this case the scalar potential for the neutral scalars may be written as
\begin{displaymath}
V=V_{D}+V_{F}+V_{\mathrm{soft}},
\end{displaymath}
where
\begin{equation}\label{eq:1dterm}
V_{D}=\frac{1}{8}(g^{2}+g'^{2})\left( |H_{u}^{0}|^{2}-|H_{d}^{0}|^{2}-|\tilde{\nu}|^{2} \right)^{2},
\end{equation}
\begin{multline}\label{eq:1fterm}
V_{F}=|\lambda_{H}H_{d}^{0}H_{u}^{0}+\frac{1}{2}\lambda_{N}\tilde{N}^{2}+\frac{1}{2}\lambda_{\Phi}\Phi^{2}|^{2}+|h^{\nu}H_{u}^{0}\tilde{N}|^{2}+|\lambda_{H}H_{u}^{0}\Phi|^{2}\\
+|\lambda_{N}\tilde{N}\Phi +h^{\nu}H_{u}^{0}\tilde{\nu} |^{2}+|h^{\nu}\tilde{\nu}\tilde{N}+\lambda_{H}H_{d}^{0}\Phi |^{2},
\end{multline}
and
\begin{multline}\label{eq:1softterms}
V_{\mathrm{soft}}=m_{H_{u}}^{2}|H_{u}^{0}|^{2}+m_{H_{d}}^{2}|H_{d}^{0}|^{2}+m_{\tilde{L}}^{2}|\tilde{\nu}|^{2}+m_{N}^{2}|\tilde{N}|^{2}+m_{\Phi}^{2}|\Phi |^{2}+\left[ A_{\nu}H_{u}^{0}\tilde{\nu}\tilde{N}\right. \\
\left. +A_{H}H_{d}^{0}H_{u}^{0}\Phi +\frac{1}{2}A_{N}\tilde{N}^{2}\Phi + \frac{1}{6}A_{\Phi}\Phi^{3}+\xi^{3}\Phi+\mathrm{h.c.}\right].
\end{multline}
The tadpole term is linear in $\Phi$ so that $\langle \Phi\rangle \neq 0$, and a $\mu$-term is always generated. 

When the electroweak symmetry is broken, $\langle H_u^0\rangle =v_u/\sqrt{2}$, $\langle H_d^0\rangle =v_d/\sqrt{2}$, either both or neither of the sneutrinos will get a VEV, since if one of them will have a VEV, there will be linear terms in $\tilde{\nu}$ and $\tilde{N}$. The minimization conditions are given in the Appendix. Neglecting the quartic term for $\tilde{\nu}$, we get an equation for the left-handed sneutrino VEV
\begin{equation}
v_{\nu}=-\frac{h^{\nu}(\lambda_{N}v_{N}v_{u}v_{\Phi}+\lambda_{H}v_{N}v_{d}v_{\Phi}+\sqrt{2}a_{\nu}v_{u}v_{N})}{m_{Z}^{2}\cos 2\beta + 2m_{\tilde{L}}^{2}},
\end{equation}
where terms proportional to $(h^{\nu})^{2}$ have been neglected in the denominator and $a_{\nu}\equiv A_{\nu}/h^{\nu}$ has been def\mbox{}ined. The expression in parentheses in the numerator is of the order of TeV$^{3}$ if we assume $a_{\nu}$, $M_{SUSY}$, and the singlet VEVs to be around a TeV. 
The denominator is twice a soft tree-level mass squared, of the order of TeV$^{2}$. 

It is known that large $A$-terms can lead to a charge or color breaking minimum in the MSSM \cite{Frere:1983ag}. The deepest minimum of the potential is achieved when the f\mbox{}ield with the smallest Yukawa coupling acquires a VEV \cite{Kitano:1999qb}.
In order to avoid a charged or colored vacuum, we require that the VEV is generated for a neutral field, in our model for a sneutrino.
Thus, the Yukawa coupling $h^{\nu}$ must be smaller than $h^{e}$. Even if $\tan \beta$ is large, using $m_e=511$ keV we need $h^{\nu}<10^{-4}$, so $\langle \tilde{\nu}\rangle$ is at most around 100 MeV. With moderate values of $\tan \beta$, values of $h^{\nu}<10^{-5}$ and $\langle \tilde{\nu}\rangle< 10$~MeV are required. Hence $v_{\nu}$ will be smaller than all other VEVs. There will be stricter constraints to $v_{\nu}$ from neutrino masses, as discussed later.

\subsection{Scalar masses at tree-level}

In the mass-squared matrix for CP-even scalars the left-handed sneutrino practically decouples from the other scalars as all its mixing terms are proportional to $h^{\nu}$ or $A_{\nu}$. 
The other four CP-even scalars may have large mixings.
Minimizing the tree-level potential gives five conditions, which can be used to eliminate the soft scalar masses $m_i^2$, with $i=H_u, \,H_d,\, \tilde L ,\,\tilde{N},\, \Phi $.
We have given the conditions in the Appendix.
For the nondecoupled fields, we use the basis $(H_{u}^{0}, H_{d}^{0}, \Phi, \tilde{N})$ and neglect terms proportional to $h^{\nu}$, $A_{\nu}$ or $v_{\nu}$, since they are small in all entries of the mass matrix. 
The mass matrices for CP-odd scalars are given in the Appendix.

The $4\times 4$ -mass matrix for CP-even neutral scalars is given by
\begin{eqnarray}
m_{11}^{2} & = & m_{Z}^{2}\sin^{2}\beta-\frac{1}{2}(A_{H}v_{\Phi}/\sqrt{2}+\lambda^{2}V^{2})\cot \beta,\label{eq:cpeven11}\\
m_{22}^{2} & = & m_{Z}^{2}\cos^{2}\beta-\frac{1}{2}(A_{H}v_{\Phi}/\sqrt{2}+\lambda^{2}V^{2})\tan \beta,\label{eq:cpeven22}\\
m_{33}^{2} & = & -\frac{\overline{\xi}^{3}}{v_{\Phi}/\sqrt{2}}+\frac{1}{2}\lambda_{\Phi}^{2}v_{\Phi}^{2}+\frac{1}{2\sqrt{2}}A_{\Phi}v_{\Phi},\\
m_{44}^{2} & = & \frac{1}{2}\lambda_{N}^{2}v_{N}^{2},\label{eq:cpeven44}\\
m_{12}^{2} & = & \frac{1}{2}(A_{H}v_{\Phi}/\sqrt{2}+\lambda^{2}V^{2})-\frac{1}{2}(m_{Z}^{2}-\lambda_{H}^{2}v^{2})\sin 2\beta,\\
m_{13}^{2} & = & \frac{1}{2}A_{H}\frac{v_{d}}{\sqrt{2}}+\lambda_{H}^{2}v_{u}v_{\Phi}+\lambda_{H}\lambda_{\Phi}v_{d}v_{\Phi}\label{eq:cpeven13},\\
m_{14}^{2} & = & \frac{1}{2}\lambda_{H}\lambda_{N}v_{d}v_{N},\label{eq:cpeven14}\\
m_{23}^{2} & = & \frac{1}{2}A_{H}\frac{v_{u}}{\sqrt{2}}+\lambda_{H}\lambda_{\Phi}v_{u}v_{\Phi}+\lambda_{H}^{2}v_{d}v_{\Phi}\label{eq:cpeven23},\\
m_{24}^{2} & = & \frac{1}{2}\lambda_{H}\lambda_{N}v_{u}v_{N},\label{eq:cpeven24}\\
m_{34}^{2} & = & \frac{1}{2}A_{N}\frac{v_{N}}{\sqrt{2}}+\frac{1}{2}\lambda_{N}\lambda_{\Phi}v_{N}v_{\Phi}+\lambda_{N}^{2}v_{N}v_{\Phi}.
\end{eqnarray}
The notations $v^2=v_{u}^{2}+v_{d}^{2}=(246\;\mathrm{GeV})^{2}$, $\lambda^{2}V^{2}=\frac{1}{2}\lambda_{H}(\lambda_{N}v_{N}^{2}+\lambda_{\Phi}v_{\Phi}^{2})$ and $\overline{\xi}^{3}=\xi^{3}+\frac{1}{4}A_{N}v_{N}^{2}+\frac{1}{2}A_{H}v_{u}v_{d}$ have been used.
All of the parameters have been assumed real for simplicity.
From the mass matrix, Eqs.~(\ref{eq:cpeven11})-(\ref{eq:cpeven22}) we see that negative $A_H$-term is preferred to get a positive-def\mbox{}inite mass matrix.
We will use in our numerical calculations one common $A$, multiplied by the corresponding coupling, and we will assume that $A$ is negative.

One can use the minimization conditions (\ref{eq:humin})-(\ref{eq:lsneutrinomin}) together with the potential, Eqs.~(\ref{eq:1dterm})-(\ref{eq:1softterms}) to f\mbox{}ind whether the R-parity breaking minimum is below the R-parity conserving minimum after EWSB. At tree-level this leads to the condition 
\begin{equation}
\frac{1}{8}\lambda_{N}^{2}v_{N}^{2}+\frac{1}{4}(\lambda_{\Phi}\lambda_{N}+2\lambda_{N}^{2})v_{\Phi}^{2}+\frac{1}{4}\lambda_{H}\lambda_{N}v^{2}\sin 2\beta+\frac{1}{\sqrt{2}}A_{N}v_{\Phi}>0.\label{eq:rpvcond}
\end{equation}
The R-parity violating minimum is below the R-parity conserving minimum in a large part of the parameter space even when $A$-terms are negative. 
In fact, if the couplings and VEVs are assumed positive, the value of $A_{N}$ must be negative 
in order to make the diagonal element of Eq. (\ref{eq:cpodd44}) in the CP-odd mass matrix positive.

The scalar sector is essentially the MSSM Higgs sector with additional two singlets. 
It has the NMSSM Higgs structure as a subset of the full scalar sector, but it is not possible to saturate the NMSSM limit \cite{Drees:1988fc}, computed from the $2\times 2$ matrix of $H_{u}$ and $H_{d}$, for the lightest scalar mass. In the NMSSM the limit is saturated by decoupling the singlet from the doublets. One can decouple $\Phi$ from the doublets by making it heavy choosing $|\xi|$  large and choosing $A_{H}$ so that the mixing terms (\ref{eq:cpeven13}) and (\ref{eq:cpeven23}) are small. $\tilde{N}$ cannot be decoupled this way in the R-parity violating case since making $\lambda_{N}^{2}v_{N}^{2}/2$ large makes also the mixing terms (\ref{eq:cpeven14}) and (\ref{eq:cpeven24}) large.

Assuming $\Phi$ is decoupled, we take the $3\times 3$ mass matrix without the elements involving $\Phi$ and then compute the determinants of $m^{2}_{3\times 3}$ and $m^{2}_{3\times 3}-m_{Z}^{2}$. One gets
\begin{eqnarray}
\mathrm{Det}(m^{2}_{3\times 3}) & = & m_{Z}^{2}\lambda_{N}^{2}v_{N}^{2}\bigg[\lambda_{H}^{2}v^{2}(\sin^{2}\beta \cos^{2}\beta -1/4)\nonumber \\
& &+\frac{1}{2}(A_{H}v_{\Phi}/\sqrt{2}+\lambda^{2}V^{2})\cot 2\beta \cos 2\beta\bigg]\label{eq:1stdet}\\
\mathrm{Det}(m^{2}_{3\times 3}-m_{Z}^{2}) & = & m_{Z}^{2}(\lambda_{N}^{2}v_{N}^{2}+\lambda_{H}^{2}v^{2}-2m_{Z}^{2})
\bigg[\lambda_{H}^{2}v^{2}\sin^{2}\beta \cos^{2}\beta\nonumber \\
& &-(A_{H}v_{\Phi}/\sqrt{2}+\lambda^{2}V^{2})\sin\beta \cos\beta\bigg]\label{eq:2nddet}.
\end{eqnarray}
Eq.~(\ref{eq:1stdet}) is a product of three masses squared, and thus needs to be positive in a stable vacuum.
If one wishes to have the lightest scalar mass heavier than $m_{Z}$, one must have $\lambda_{N}^{2}v_{N}^{2}>2m_{Z}^{2}$ (see eq. (\ref{eq:cpeven44})), so in order to have the lightest mass above $m_{Z}$, both of the expressions in square brackets
in Eqs.~(\ref{eq:1stdet})-(\ref{eq:2nddet}) need to be positive. 
However, they cannot be made positive simultaneously, so at least one eigenvalue is below $m_{Z}$, if the vacuum is stable. 
It is easy to saturate this limit by taking\footnote{In addition all the diagonal elements must be greater or equal to $m_{Z}^{2}$.} $\tan \beta \gg 1$ so that the right-hand side in (\ref{eq:2nddet}) becomes zero and thus $m_{Z}^{2}$ is an eigenvalue. Therefore the lightest scalar mass is constrained similarly than in the MSSM.
If Eq. (\ref{eq:2nddet}) is chosen to be positive, the vacuum is not stable. This indicates that if one tries to push the lightest scalar mass above $m_{Z}$ one arrives at an R-parity conserving vacuum. If one assumes R-parity conservation, the sneutrino state decouples and hence the tree-level limit is the same as in the NMSSM, which can be clearly above $m_{Z}$.

Similarly than in the MSSM, one could rely in the spontaneously R-parity violating NMSSM (SRPV-NMSSM) on the large radiative corrections on the scalar masses to achieve a Higgs boson with $m_H=$125 GeV.
Alternatively, one or two of the lightest scalars can be mainly singlet and the $125$~GeV Higgs boson is a heavier scalar, which may have a tree-level mass above $m_{Z}$. 
Unless the soft tadpole term is small, the lighter than 125 GeV scalar is sneutrino-like, and the other singlet dominated Higgs is rather heavy.

The mass of the SM-like ($H_{u}$-dominated) Higgs is usually not much above $m_{Z}$ at tree-level. The only exception is the combination of large $\lambda_{H}$, small $v_{N}$ and $\tan \beta \rightarrow 1$. As can be seen from Eq. (\ref{eq:1stdet}) the tree-level mass of the lightest CP-even scalar goes to zero in this limit. 
With small values of $v_{N}$ the sneutrino-like state remains lighter than the lighter doublet state and will be the one whose mass tends to zero. 
In that case for large $\lambda_H$ the SM-like Higgs mass can be lifted very much like in the R-parity conserving (RPC) NMSSM.

Since the eigenvalue equation is of the third (or fourth) order, the analytical form of the eigenvalues is not illuminating. The expression may be simplif\mbox{}ied only in certain limiting cases. 
For illustration we shall compute one case. We shall take new combinations of the doublet Higgses, $h=\sin\beta H_{u}^{0}+\cos\beta H_{d}^{0}$ and $H=\cos\beta H_{u}^{0}-\sin\beta H_{d}^{0}$. The $h$-state has the same VEV than the SM Higgs and hence also the same couplings. We shall look at the mixing between $h$ and $\tilde{N}$ in the case when $H$ is so heavy that it decouples. The tree-level mass matrix in the basis $(h,H,\tilde{N})$ is
\begin{equation}\label{eq:hHNmatrix}
\begin{pmatrix}
m_{Z}^{2}+\lambda_{H}^{2}v^{2}\sin^{2}2\beta & -\frac{1}{2}(m_{Z}^{2}-\lambda_{H}^{2}v^{2}/2)\sin 4\beta & \lambda_{H}\lambda_{N}v_{N}v\sin 2\beta\\
-\frac{1}{2}(m_{Z}^{2}-\lambda_{H}^{2}v^{2}/2)\sin 4\beta & -(A_{H}\frac{v_{\Phi}}{\sqrt{2}}+\lambda^{2}V^{2})\frac{\tan\beta +\cot\beta}{2}-\lambda_{H}^{2}v^{2}\sin^{2}2\beta & \lambda_{H}\lambda_{N}v_{N}v\cos 2\beta\\
\lambda_{H}\lambda_{N}v_{N}v\sin 2\beta & \lambda_{H}\lambda_{N}v_{N}v\cos 2\beta & \frac{1}{2}\lambda_{N}^{2}v_{N}^{2}
\end{pmatrix}
\end{equation}
We look at the $2\times 2$ submatrix formed by $h$ and $\tilde{N}$, whose eigenvalues can be solved exactly. The eigenvalues are
\begin{equation}\label{eq:apprmass}
\frac{1}{2}(m_{Z}^{2}+\lambda_{H}^{2}v^{2}\sin^{2}2\beta +\lambda_{N}^{2}v_{N}^{2}/2)\pm \sqrt{\frac{1}{4}(m_{Z}^{2}+\lambda_{H}^{2}v^{2}\sin^{2}2\beta -\lambda_{N}^{2}v_{N}^{2}/2)^{2}+ \lambda_{H}^{2}\lambda_{N}^{2}v^{2}v_{N}^{2}\sin^{2}2\beta}.
\end{equation}
If $\lambda_{N}^{2}v_{N}^{2}/2<m_{Z}^{2}+\lambda_{H}^{2}v^{2}\sin^{2}2\beta$ and we assume the last term inside the square root to be small (e.g. in the limit of large $\tan \beta$) compared to $\left( m_{Z}^{2}+\lambda_{H}^{2}v^{2}\sin^{2}2\beta -\lambda_{N}^{2}v_{N}^{2}/2\right)^2$ we can expand the square root and get
\bea
m_{\tilde{N}}^{2} & = & \frac{1}{2}\lambda_{N}^{2}v_{N}^{2}-\frac{\lambda_{H}^{2}\lambda_{N}^{2}v^{2}v_{N}^{2}}{m_{Z}^{2}+\lambda_{H}^{2}v^{2}\sin^{2}2\beta -\lambda_{N}^{2}v_{N}^{2}/2}\sin^{2}2\beta,\\
m_{H}^{2} & = & m_{Z}^{2}+\lambda_{H}^{2}v^{2}\sin^{2}2\beta + \frac{\lambda_{H}^{2}\lambda_{N}^{2}v^{2}v_{N}^{2}}{m_{Z}^{2}+\lambda_{H}^{2}v^{2}\sin^{2}2\beta -\lambda_{N}^{2}v_{N}^{2}/2}\sin^{2}2\beta.
\eea
From these expressions one sees that it is possible to get a Higgs-like state heavier than $m_{Z}$.

The approximations leading to Eq. (\ref{eq:apprmass}) are valid if the $H$-state can be decoupled from the other states. This is the case when either $\tan \beta$ or $\cot \beta$ is large.
The tree-level mass of the second lightest scalar increases at values of $\tan \beta$ close to one, but in that limit the approximations done here are not reliable.

The CP-odd scalar mass matrix is given in the Appendix.
Among the CP-odd scalars there is one Goldstone with the composition $\cos \beta \mathrm{Im} (H_{d}^{0})-\sin \beta \mathrm{Im}(H_{u}^{0})$, as expected. In the limit $\lambda_{N}, A_{N}\rightarrow 0$ with $v_{N}\neq 0$ also the CP-odd component of $\tilde{N}$ becomes massless, since it is the pseudo-Majoron, as explicitly demonstrated in \cite{Frank:2007un}.

If two more generations are included in the model, 
the dif\mbox{}ference to the above is that all mass matrices are larger in size and there are more neutral scalars that acquire VEVs. However one can always choose such a linear combination of the singlet sneutrinos that only one of them has a VEV. This combination is very close to one of the physical states, since all the mixing terms with other neutral scalars for the combinations without a VEV include  $h^{\nu}$. Thus the essential features of the full model are similar to the model with one generation. 
All three left-handed sneutrinos decouple from the rest of the scalar sector and the mixing at tree-level between the doublet scalars and singlet sneutrino combinations without a VEV comes only through the tiny neutrino Yukawa term.
The singlet sneutrinos which do not get a VEV will get their masses from the soft SUSY breaking terms and from the VEVs of other scalars via the generalization of the first terms of equation (\ref{eq:1fterm}).

\subsection{Constraints from fermion masses}

Neutrino mass generation in a similar model but with three neutrino generations was considered in \cite{Kitano:1999qb, Frank:2005tn, Frank:2007un}.
It was found that mass differences and mixing angles compatible with experimental results can be generated. 
The tree-level mass matrix for neutral fermions with one generation in our model is
\begin{equation}
\begin{pmatrix}
0 & h^{\nu}v_{u}/\sqrt{2} & 0 & h^{\nu}v_{N}/\sqrt{2} & 0 & -g'v_{\nu}/\sqrt{2} & gv_{\nu}/\sqrt{2}\\
h^{\nu}v_{u}/\sqrt{2} & \lambda_{N}v_{\Phi}/2\sqrt{2} & 0 & h^{\nu}v_{\nu}/\sqrt{2} & \lambda_{N}v_{N}/\sqrt{2} & 0 & 0\\
0 & 0 & 0 & \lambda_{H}v_{\Phi}/\sqrt{2} & \lambda_{H}v_{u}/\sqrt{2} & -g'v_{d}/\sqrt{2} & gv_{d}/\sqrt{2}\\
h^{\nu}v_{N}/\sqrt{2} & h^{\nu}v_{\nu}/\sqrt{2} & \lambda_{H}v_{\Phi}/\sqrt{2} & 0 & \lambda_{H}v_{d}/\sqrt{2} & g'v_{u}/\sqrt{2} & -gv_{u}/\sqrt{2}\\
0 & \lambda_{N}v_{N}/\sqrt{2} & \lambda_{H}v_{u}/\sqrt{2} & \lambda_{H}v_{d}/\sqrt{2} & \lambda_{\Phi}v_{\Phi}/2\sqrt{2} & 0 & 0\\
-g'v_{\nu}/\sqrt{2} & 0 & -g'v_{d}/\sqrt{2} & g'v_{u}/\sqrt{2} & 0 & M_{1} & 0\\
gv_{\nu}/\sqrt{2} & 0 & gv_{d}/\sqrt{2} & -gv_{u}/\sqrt{2} & 0 & 0 & M_{2}\\
\end{pmatrix},
\end{equation}
where the basis $(\nu, N, \tilde{H}_{d}^{0}, \tilde{H}_{u}^{0}, \tilde{\Phi}, \tilde{B}^{0}, \tilde{W}^{0})$ and the convention $v_{u}^{2}+v_{d}^{2}+v_{\nu}^{2}\simeq v_{u}^{2}+v_{d}^{2}=(246$~GeV$)^{2}$ is used.

The left-handed neutrino will get a mass via the seesaw mechanism and the sneutrino VEVs according to the usual seesaw relation.
If $v_{N}\ll v_{\Phi}$ the seesaw mass from the upper left-corner is
$m_\nu = \sqrt{2}(h^{\nu}v_{u})^{2}/\lambda_{N}v_{\Phi}$.
Since $\lambda_{N}$ is the measure of explicit lepton number violation, the pseudo-Majoron mass depends on it (see \cite{Frank:2007un}). 
If the left-handed neutrino mass is $m_{\nu}\simeq 0.1$~eV, we must have $v_{\Phi}\sim (h^{\nu})^{2}\times 10^{12}$~TeV. 
On the other hand the ef\mbox{}fective $\mu$-parameter is $\lambda_{H}v_{\Phi}$ and to avoid the f\mbox{}ine-tuned cancellation with the Z-boson mass, we need to have $\lambda_{H}v_{\Phi}$ around the weak scale. If $v_{\Phi}$ is large, we need $|\lambda_{H}|\ll 1$.
We will assume here $|\lambda_{H}|$ close to one and $h_\nu \leq 10^{-6}$.
The singlet sneutrino VEV $v_N$ also contributes to the neutrino mass. 
The contribution is significant only if $v_{\Phi}\ll v_{N}$. In that limit the type-I seesaw mass matrix is of a pseudo-Dirac form and contributes little due to the small Yukawa coupling. The sneutrino VEV generates ef\mbox{}fective bilinear R-parity violating terms, which generate neutrino masses. For values of $h^{\nu}$ of the order of $10^{-6}$, viable neutrino masses require bilinear R-parity violating terms around $10$~MeV or smaller \cite{Gozdz:2008zz}, which constrain $v_{N}<10$~TeV. Thus all VEVs and therefore all the masses are at a scale of at most ten TeV's.

The order of $v_{N}$ and $v_{\Phi}$ is also relevant when inspecting the second-lightest neutral fermion.
If $v_{\Phi}\ll v_{N}$, the second lightest neutral fermion is a neutralino having also a significant singlino component. 
If $v_{N}\ll v_{\Phi}$ the second-lightest neutral fermion is mainly a mixture of the singlet neutrino and singlet higgsino. The mixture depends on the gaugino mass parameters as well. If $v_{N}$ and $v_{\Phi}$ are both large compared with $M_1$ or $M_2$, and $|\lambda_{H}|$ is small the second-lightest fermion is a neutralino with a large gaugino component.

The left-handed sneutrino VEV is constrained by the neutrino mass constraints. The terms $gv_{\nu}$ can give the neutrino a mass via the gaugino seesaw mechanism \cite{Kitano:1999qb}.
The gaugino seesaw gives the neutrino a mass of $g^{2}v_{\nu}^{2}/2M_{2}$. This constrains $v_{\nu}$ to be 1~MeV or less if we assume for the gaugino mass parameter $M_2=1$ TeV.
In our numerical calculations we will require the tree-level neutrino masses to be in the range
$0.05\,\mathrm{eV}< m_{\nu}<0.5\,\mathrm{eV}$.

The charged fermion mass matrix is of the form
\begin{equation}
\begin{pmatrix}
\ell^{+}_{R} & \tilde{H}_{u}^{+} & \tilde{W}^{+}
\end{pmatrix}
\mathcal{M}
\begin{pmatrix}
\ell^{-}_{L} \\ 
\tilde{H}_{d}^{-}\\
\tilde{W}^{-}
\end{pmatrix}
\end{equation}
where the mass matrix is
\begin{equation}\label{eq:leptonmass}
\mathcal{M}=
\begin{pmatrix}
h^{e}v_{d} & -h^{\nu}v_{N} & 0\\
-h^{e}v_{\nu} & -\lambda_{H}v_{\Phi} & -gv_{u}\\
-gv_{\nu} & -gv_{d} & M_{2}
\end{pmatrix}.
\end{equation}
The experimental lower limit on the chargino mass, when R-parity is conserved, is around 103 GeV.
This may change when R-parity is broken.
Assuming pure wino and leptonic R-parity breaking, and only one dominant coupling of $LLE$ type, the limit becomes around 540 GeV for 
neutralinos lighter than the wino and heavier than $300$~GeV, and around 500 GeV for neutralinos between 100 and 300 GeV \cite{ATLAS:2012kr}. One has to note that usually the experimental bounds are derived assuming a single RPV coupling to dominate. In SRPV there are both LQD- and LLE-type operators which can make the signal and the limits dif\mbox{}ferent from that of a single RPV coupling.
In our numerical calculations we will require that the ef\mbox{}fective $\mu$-parameter is $\ge 103$ GeV,
and we choose for our gaugino mass parameters $M_1$=300 GeV, $M_2$=600 GeV, and 
$M_3$=1.5 TeV.

\section{Numerical results}

We consider the  model with one generation in detail, since the two singlet sneutrinos which do not have VEVs will not mix with the other neutral scalars. Therefore, the results will be similar for three generations.

We calculate the Higgs boson mass in the model with the ef\mbox{}fective potential approach including one-loop corrections from top-stop and bottom-sbottom loops. We then perform a random scan over the model parameters.  
The minimum and maximum values for the parameters in the random scan are given in Table \ref{tb:scanvalues}. The values for the trilinear couplings are computed by multiplying the corresponding coef\mbox{}f\mbox{}icient in the superpotential by a common parameter $A$, 
{\it e.g.} $A_{N}=A\lambda_{N}$.
Values of $|A|$ less than $1$~TeV do not produce any RPV minima. 
As discussed before, negative $A$ is preferred.
Also the sign of $\xi$ needs to be negative if $v_{\Phi}$ is chosen positive, because $\partial V/\partial \Phi =\xi^{3}$ when all f\mbox{}ields have zero values and hence a negative value of $\xi$ implies a minimum for positive $\langle \Phi\rangle$.
\begin{table}
\begin{center}
\begin{tabular}{|c|c|c|}
\hline
Parameter & Minimum value & Maximum value\\
\hline
$\tan \beta$ & 1 & 30\\
$\lambda_{N}$ & 0.1 & 0.8\\
$\lambda_{\Phi}$ & 0.05 & 0.25\\
$\lambda_{H}$ & 0.1 & 0.6\\
$h_{\nu}$ & 10$^{-8}$ & 10$^{-6}$\\
$A$ & -6000 & -1000\\
$\xi$ & -3500 & -500\\
$v_{\Phi}$ & 600 & 2500\\
$v_{N}$ & 50 & 4000\\
$v_{\nu}$ & 10$^{-6}$ & 5$\cdot$10$^{-4}$\\
$m_{\tilde{Q}_{L},\tilde{t}_{R},\tilde{b}_{R}}$ & 700 & 1700\\
\hline
\end{tabular}
\end{center}
\caption{The minimum and maximum values for the parameters of the random scan. Parameters which have dimensions are given in GeV's. The gaugino mass parameters are f\mbox{}ixed and chosen to be $M_{1}=300$~GeV, $M_{2}=600$~GeV and $M_{3}=1.5$~TeV. \label{tb:scanvalues}}
\end{table}
The minima of the ranges for soft squark masses are chosen to give stop and sbottom masses which are compatible with RPV squark searches \cite{Chatrchyan:2013xsw}.

The condition for an RPV minimum, Eq. (\ref{eq:rpvcond}), with $A_{N}$ negative, implies that either $v_{N}$ or $v_{\Phi}$ must be sizable to have an RPV vacuum. These parameters in turn largely determine the singlet-dominated scalar masses. 
Hence typically at least one of the singlets is heavy. 
If the sneutrino-like state is heavy, we have the NMSSM but with a more stringent tree-level bound on the lightest scalar mass as discussed earlier.
We choose the parameters so that the NMSSM-like singlet is usually heavy and let the sneutrino mass vary over a larger range. We will later comment on enlarging the range for $v_\Phi$.

We check whether the potential gets a value below the symmetric vacuum ($v_{i}=0$ for all f\mbox{}ields) when $\Phi$ is given a VEV, $\Phi$, $H_{u}^{0}$ and $H_{d}^{0}$ are given a VEV or when all of the scalar f\mbox{}ields develop VEVs. When $\xi<0$ almost always $v_{\Phi}\neq 0$ produces a vacuum with a smaller energy than with $v_{\Phi}=0$. With our choices of parameters spontaneous R-parity violation produces a lower vacuum energy in more than 85 \% of the data points. With $500 000$ initial points the number of points satisfying each criterion are given in Table \ref{tb:cutflow1}.

\begin{table}
\begin{center}
\begin{tabular}{|l|l|}
\hline
Total number of points & 500 000\\
Mu-term generated ($v_{\Phi}\neq 0$) & 484 397\\
EW symmetry broken ($v_{u}$, $v_{d}\neq 0$) & 457 508\\
R-parity broken ($v_{N}$, $v_{\nu}\neq 0$) & 438 802\\
Neutrino mass within limits & 90 303\\
A CP-even scalar within LHC mass limits & \\
and LEP limits for the lightest Higgs & 13 750\\
\hline
\end{tabular}
\end{center}
\caption{The number of points satisfying the criteria listed. The tree-level neutrino mass is required to be between 0.05--0.5~eV and the CP-even scalar mass between 120--130~GeV. \label{tb:cutflow1}
Each of the listed event numbers satisfy also the limits mentioned before.}
\end{table}

The parameter ranges are chosen so that the probability to have an R-parity violating vacuum satisfying the other constraints would be high. We made also some other scans with less points satisfying the criteria. 
In particular, we took smaller values of $v_{\Phi}$ ($100\ldots 600$~GeV) and got about one third of the points compared to the main data set. One reason is that for $v_{\Phi}< 200$~GeV the bound on the chargino mass is very restrictive.
Qualitatively the biggest change in the spectrum is that the lightest CP-odd scalar becomes lighter, to around one third of the value in our main data set.
This will result in more points discarded by $B_s\rightarrow \mu\mu $ constraint than in the case of the 
main data set, to be discussed later.

\subsection{Neutral scalar sector}

\begin{figure}
\begin{center}
\includegraphics[width=\textwidth]{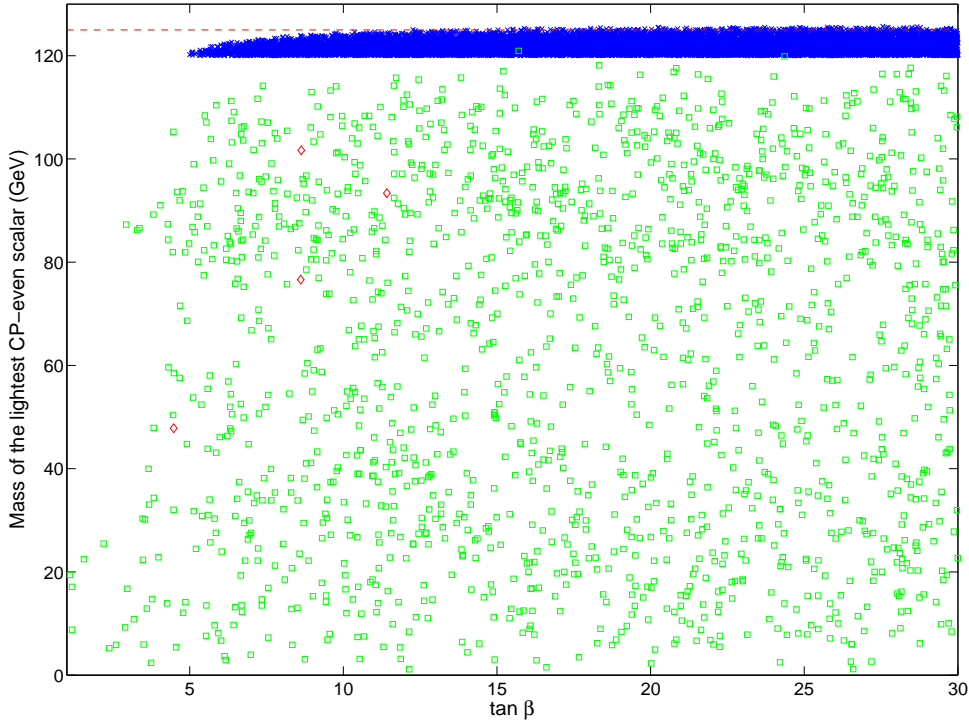}
\end{center}
\caption{The masses of the lightest CP-even scalar as a function of $\tan \beta$ using the one-loop ef\mbox{}fective potential. The largest component of the lightest scalar is $H_{u}$ (blue crosses, 12217 points), $H_{d}$ (red diamonds, 4 points) or $\tilde{N}$ (green squares, 1531 points). The red dashed line is at $125$~GeV. These points have an R-parity violating stable vacuum and satisfy the experimental constraints on the Higgs masses from LEP and LHC.
\label{fig:higgsmass}}
\end{figure}

Taking into account the experimental limits for the Higgs sector, {\it i.e.} a scalar with $m_h\sim 125$ GeV and 
the other scalars satisfying LEP and LHC limits, the distribution of the lightest CP-even scalar is given in Figure \ref{fig:higgsmass}.
%
We assume that the $125$~GeV particle found by LHC is a CP-even scalar that is responsible for electroweak symmetry breaking. 
With $\tan \beta>1$ the dominant component of such a scalar is $H_{u}^{0}$. 
This scalar can be the lightest CP-even scalar in our model. 
Since we have computed the mass matrices only to one-loop accuracy and restricted ourselves to third sector squark-quark corrections\footnote{The right-handed neutrino-sneutrino contribution to the Higgs mass can be a few GeVs \cite{Wang:2013jya}.}, we accept all such points where the lightest CP-even scalar has a mass between $120$~GeV and $130$~GeV.

The second option is that the lightest CP-even scalar is below $120$~GeV but it is mainly a singlet so that the LEP limits for the $hZZ$-coupling \cite{Barate:2003sz} are satisfied\footnote{The constraints we use for the lightest CP-even scalar are that the doublet component must be below $4 \%$ if $m < 80$~GeV and below $25 \%$ if $80$~GeV $< m <120$~GeV.}
The second lightest CP-even scalar in our model can then be mainly $H_{u}^{0}$ and in the mass range between $120$~GeV and $130$~GeV. 
The mass of the SM-like Higgs can be quite large, up to $135$~GeV if the lightest scalar is between $80$ and $120$~GeV. 
With our choice of parameters the particle that can be lighter than the SM-like Higgs is usually mostly a sneutrino.
It is also possible that there is another scalar below 120 GeV, which is mainly a singlet.
In our data set there was a single data point with two light singlets. 
In the data set with $v_{\Phi}$ ($100\ldots 600$~GeV) the sneutrino dominated scalar remains as the lighter singlet, but the portion of $\Phi$ singlet in the lightest scalar can rise to about 5 \%, whereas it is negligible with larger values of $v_{\Phi}$.

\begin{figure}
\begin{center}
\includegraphics[width=0.85\textwidth]{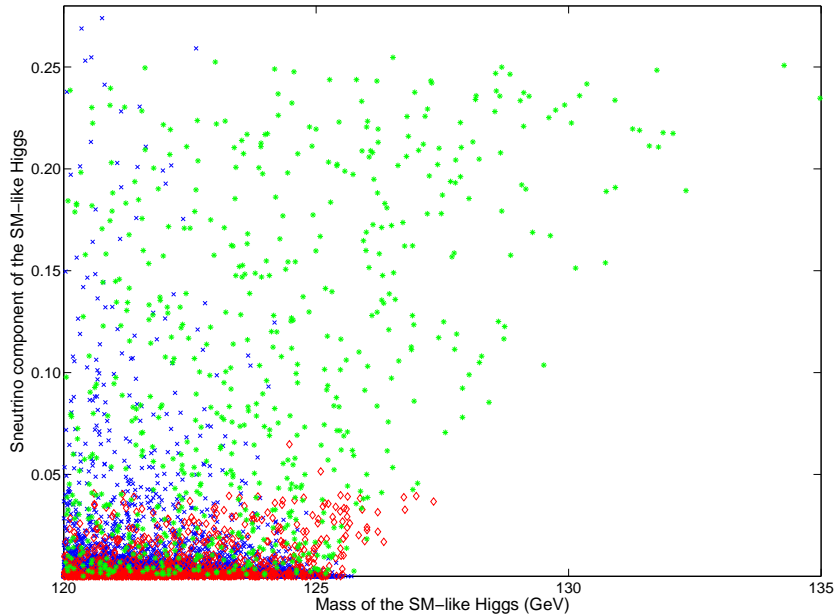}
\end{center}
\caption{The sneutrino component of the SM-like Higgs as a function of the Higgs mass. 
Blue crosses are points where the SM-like Higgs is the lightest scalar, red diamonds have a lighter scalar below $80$~GeV and green asterisks have a lighter scalar between $80$ and $120$~GeV. 
The highest masses require relatively large Higgs-sneutrino mixing. \label{fig:sneutmix}}
\end{figure}

In the case of NMSSM the Higgs mass can be lifted due to the mixing between the doublets and the singlet $\Phi$ \cite{Badziak:2013bda}.
In the model with spontaneous R-parity violation, the doublet-sneutrino mixing is essential in lifting the SM-like Higgs mass, even by some 8 GeV. 
In Figure \ref{fig:sneutmix} we plot the sneutrino component of the SM-like Higgs as a function of the Higgs mass. The largest Higgs masses require signif\mbox{}icant Higgs-sneutrino mixing.

\begin{figure}
\begin{center}
\includegraphics[width=0.85\textwidth]{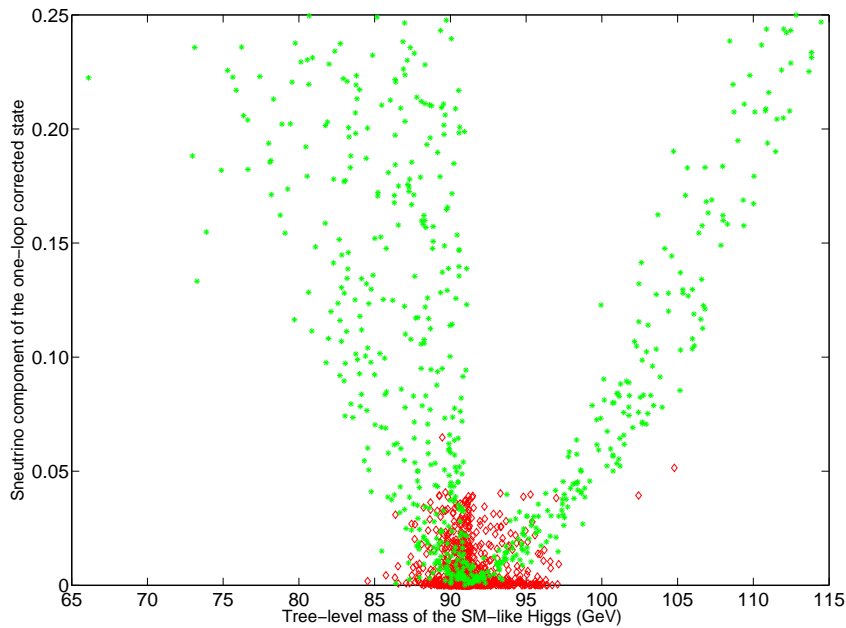}
\end{center}
\caption{The sneutrino component of the SM-like Higgs as a function of its tree-level mass.  The color coding is the same as in Figure \ref{fig:sneutmix}. See the text for explanation. \label{fig:treemix}}
\end{figure}

To estimate the size of the mixing ef\mbox{}fect to the masses we show in Figure \ref{fig:treemix} the relationship between the tree-level masses and Higgs-sneutrino mixing. We take the sneutrino components from the one-loop corrected mass matrix so that comparison to Figure \ref{fig:sneutmix} is easy. The interpretation of the plot is not straightforward. If the sneutrino-dominated state at one-loop is below $80$~GeV the mixing can lift the SM-like Higgs mass to around $95$~GeV. This is compatible with the results at one-loop level.
In the mass matrix, Eq. (\ref{eq:hHNmatrix}), the top-stop loop corrections increase the value of the $(1,1)$-element but do not af\mbox{}fect the $(3,3)$-element. Hence at tree-level the $(1,1)$-element can be smaller than the $(3,3)$-element even if the opposite is true at one-loop level.
If at tree-level the SM-like Higgs is lighter than the sneutrino-like state, the mixing brings the mass of the Higgs-like state down instead of lifting it, which can be seen in Figure \ref{fig:treemix}. At tree-level the diagonal elements of the mass matrix are closer to each other than when loop corrections are taken into account. This tends to overshoot the mixing ef\mbox{}fect, which is why it looks as if the masses can increase by more than $20$~GeV by mixing, whereas the actual enhancement of one-loop masses is less than $10$~GeV.

\begin{figure}
\begin{center}
\includegraphics[width=0.85\textwidth]{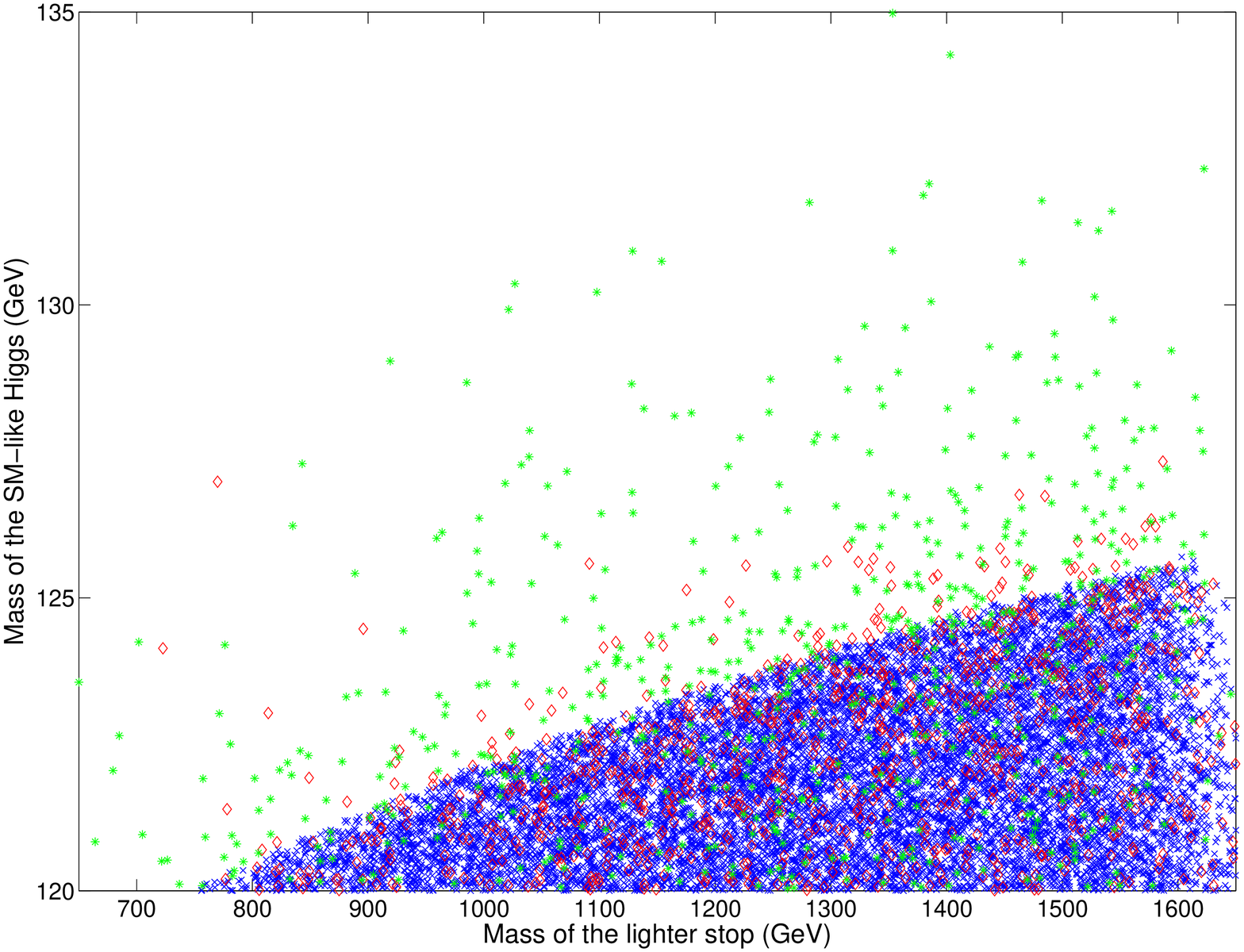}
\includegraphics[width=0.85\textwidth]{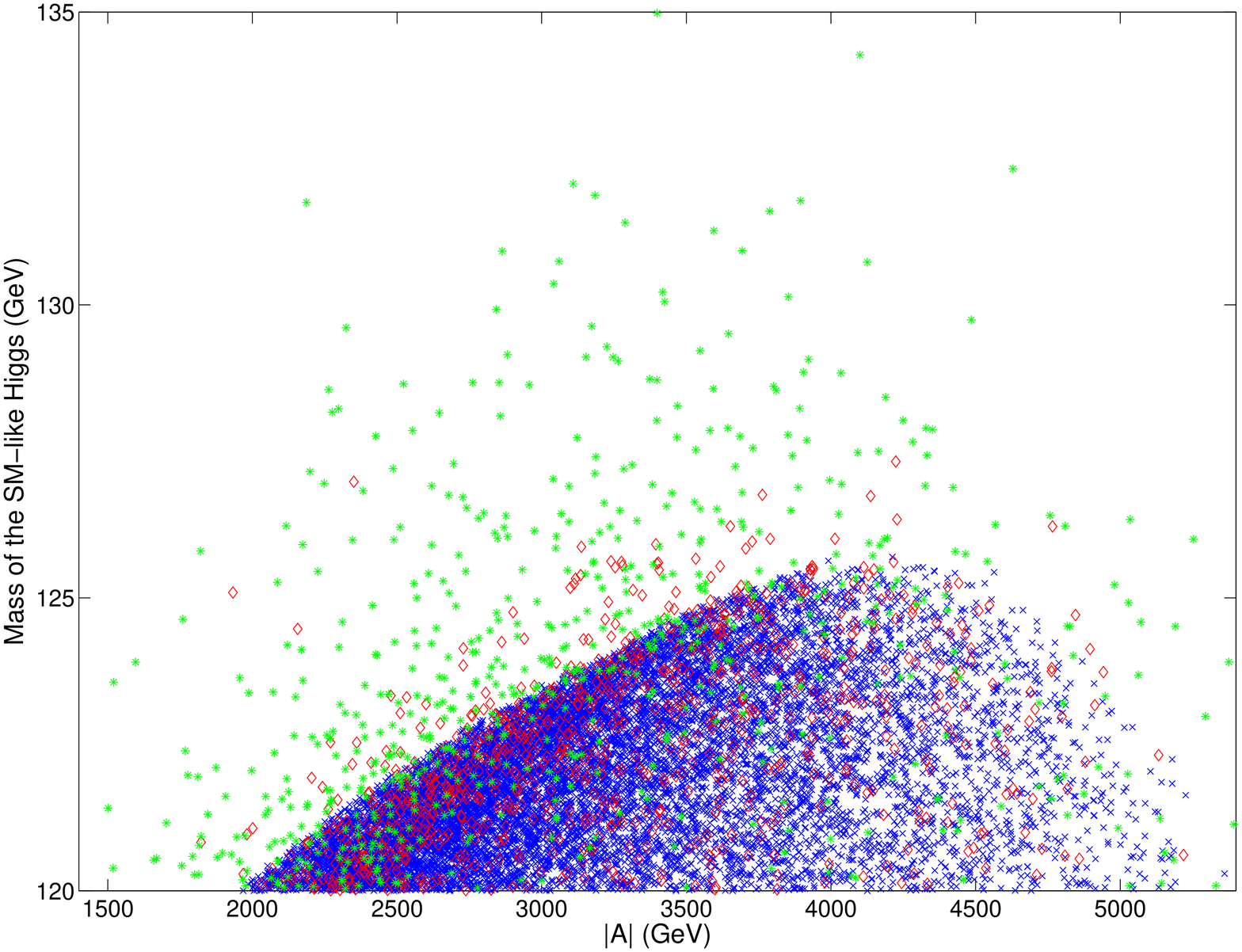}
\end{center}
\caption{The mass of the SM-like Higgs as a function of the lighter stop mass and $|A|$. The upper mass limit of the SM-like Higgs has not been applied in these plots.
The color coding is same as in Figure \ref{fig:sneutmix}.\label{fig:sneut_mix_v2}
\label{fig:mhvstop}}
\end{figure}

It is well known that in the MSSM it is necessary to have either a heavy stop, $m_{\tilde t}\sim {\cal{O}}$(1 TeV),
or large mixing between the stops, in order to lift the Higgs mass to the measured value.
When the SM-like Higgs is the lightest scalar the situation is similar to the MSSM. 
However, if there is a lighter scalar than the 125 GeV Higgs the stops can be significantly lighter than in the MSSM. 
With our choice of parameters the stop masses are dominated by the soft SUSY breaking masses. 
The SM-like Higgs mass has been plotted as a function of the lighter stop mass in Figure \ref{fig:mhvstop}.
From Fig.~\ref{fig:mhvstop} we note that the observed Higgs mass can be found with stops well below  $1$~TeV.
The lighter stop usually has a mass somewhere between $700$~GeV and $1500$~GeV. 
In our data set there are points where a $125$~GeV Higgs mass is obtained with stop mass around $700$~GeV and a $120$~GeV Higgs mass is obtained with stop mass as light as $400$~GeV.
The mixing between the stops is large due to the large $A$-terms needed for spontaneous RPV, and
the mass dif\mbox{}ference between the stops 
is typically around $200$~GeV. 

Assuming stop decay to a top quark and a neutralino\footnote{Note that with R-parity violation the decay modes may be very different, {\it e.g.} stop could be the LSP \cite{Marshall:2014kea}.} and the neutralino subsequently decaying to leptons, the latest CMS searches for stops with RPV decays give lower mass limits of the order $800$--$1000$~GeV \cite{Chatrchyan:2013xsw}. 
The upgraded LHC with $3000$~fb$^{-1}$ may reach to stop masses close to $2$~TeV in models with leptonic R-parity violation \cite{Duggan:2013yna}. Hence, with the mentioned assumption, the LHC should be able to f\mbox{}ind the stop in the region where the Higgs mass is lifted by the Higgs-sneutrino mixing.

The two subsets ($125$~GeV Higgs the lightest/second lightest) are distributed quite uniformly in the parameter space. The main dif\mbox{}ferences between the two categories are that in the case where the $125$~GeV scalar is the lightest one, $v_{N}$ and $|A|$ are larger than in the other category. 
These can be understood as follows.
If $v_{N}$ is small, the tree-level contribution from equation (\ref{eq:cpeven44}) to the sneutrino mass is small and the light scalar becomes mostly a sneutrino. 
If a light sneutrino exists in the model, the tree-level bound of $m_{Z}$ does not apply any more to the doublet Higgs mass and thus smaller $|A|$-terms can lift the Higgs mass to $\sim 125$~GeV. This is shown in the lower plot of Figure \ref{fig:mhvstop}.

In a large number of the sample points there is an almost pure singlet scalar and an almost pure sneutrino. 
In that case the doublet Higgses will look like those of the MSSM. With our choice of the range for $\xi$ the singlet $\Phi$ dominated Higgs tends to be the heaviest of the CP-even scalars. It almost never has a large mixing with the doublet Higgses but may mix with the right-handed sneutrino.
The state with the largest sneutrino component has a mass that is driven by the tree-level contribution from equation (\ref{eq:cpeven44}). Since we let the parameters $\lambda_{N}$ and $v_{N}$ vary over a wide range of values, this scalar may be light or heavy. If both $\lambda_{N}$ and $v_{N}$ are small, it is the lightest state. In most of our parameter sets, it is the second lightest after the lighter (SM-like) doublet Higgs. The large values of $|A|$ usually make the second doublet Higgs mass rise up to more than $1$~TeV.

\begin{figure}
\begin{center}
\includegraphics[width=\textwidth]{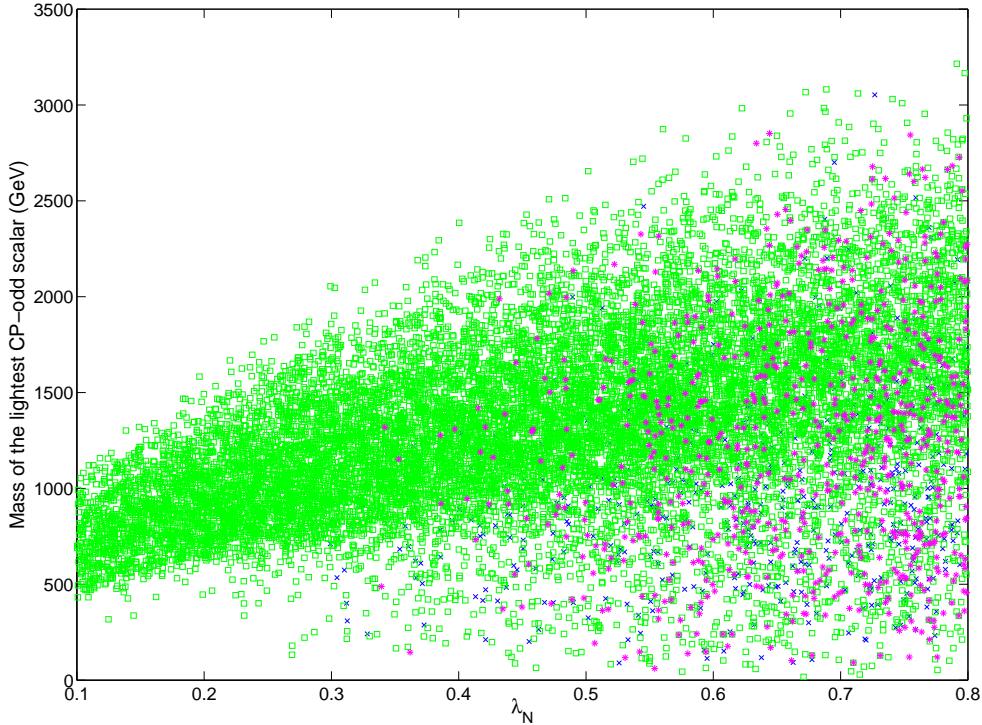}
\end{center}
\caption{The lightest CP-odd Higgs boson mass as a function of $\lambda_{N}$. Blue crosses indicate a doublet scalar, green squares a sneutrino and magenta asterisks a singlet scalar. \label{fig:ma}}
\end{figure}

The lightest CP-odd scalar mass as a function of $\lambda_{N}$ is plotted in Figure \ref{fig:ma}.
In the CP-odd sector the lightest state is usually the pseudo-Majoron. It gets its mass from the explicitly lepton-number violating terms. Hence it is essentially a mixture of the singlets (usually sneutrino-dominated) and its mass depends mostly on $\lambda_{N}$ and $v_{\Phi}$. With large values of these parameters even the lightest CP-odd scalar may be above $1$~TeV. The large values of $|A|$ make the MSSM-like CP-odd Higgs heavy, almost always more than a TeV. However at large values of $\lambda_{N}$ we have some points where the lightest CP-odd Higgs is the MSSM-like one and it can be rather light. 
We will return to the explanation of this phenomenon in Section~\ref{sec:couplings}.

In the charged scalar sector, charged Higgs doublets mix with charged sleptons.
The charged slepton mixings with charged Higgses are all suppressed by $h^{\nu}$, $v_{\nu}$ or $A_{\nu}$.
In our model, the charged Higgs mass will always be above $1$~TeV due to the large $A$-terms needed for 
spontaneous RPV.

\subsection{Constraints from f\mbox{}lavor changing rare processes}

In the model there are two sources for lepton f\mbox{}lavor violation. 
On one hand, the neutrino Yukawa couplings can be non-diagonal leading to lepton f\mbox{}lavor violation.
On the other hand, when R-parity is spontaneously violated, 
in the fermionic sector charged leptons mix with charginos, and neutrinos with neutralinos, and
correspondingly in the scalar sector all the neutral (charged) scalars, including sneutrinos (charged sleptons), mix with each other.
Thus, constraints from lepton f\mbox{}lavor violating processes can be expected.
The lepton f\mbox{}lavor violating muon decays $\mu\to e\gamma$ and $\mu \to eee$ are experimentally very constrained, with upper limits of  $\mathrm{BR}(\mu \to e\gamma)< 5.7\times 10^{-13}$ \cite{Adam:2013mnn} and $\mathrm{BR}(\mu \to eee)<1.0\times 10^{-12}$ \cite{Bellgardt:1987du}. 

Since in our model the right-handed neutrinos are at the TeV scale, they might noticeably contribute to these reactions. 
The limits for neutrino Yukawa couplings in type-I seesaw models are of the order of $10^{-2}$ for right-handed neutrino masses of $100$~GeV or more \cite{Ibarra:2011xn,Dinh:2012bp}. 
Since successful spontaneous R-parity violation requires smaller Yukawa couplings by many orders of magnitude, the right-handed neutrinos will not contribute measurably to the muon decays. 

The bounds on spontaneous R-parity violation parameters with the MSSM f\mbox{}ield content were analyzed in \cite{Frank:2001tr}. The bounds from $\mu \to e\gamma$ and muon-electron conversion are the most stringent ones. 
From the upper limits on branching ratios we deduce upper limits for charged lepton-chargino mixing.
As is found from the mass matrix (\ref{eq:leptonmass}) the charged lepton-chargino mixing is always suppressed by either $v_{\nu}/v$ or $h^{\nu}$ and is therefore small. This will suppress lepton f\mbox{}lavor violating decays coming from spontaneous RPV. 
The maximum amount of lepton-chargino mixing in the muon or electron sector is $3.25\times 10^{-5}$ in absolute value using the data set generated by the parameter scan. Hence any product of mixing elements is bound by $|V_{ij}U_{ik}|<1.05\times 10^{-9}$. Even after taking into account that the experimental bound has been tightened by a factor of 20, this is three orders of magnitude smaller than the bounds in \cite{Frank:2001tr} where it was assumed that $m_{\tilde{f}}=100$~GeV and $\tan \beta =2$. The lower bounds on sfermion masses in the R-parity violating case are around $100$~GeV, but there may be enhancement for larger values of $\tan \beta$. 
However, all the data points clearly satisfy $BR(\mu \to e\gamma)$  constraints.

%
\begin{fmffile}{higgsgraphs}
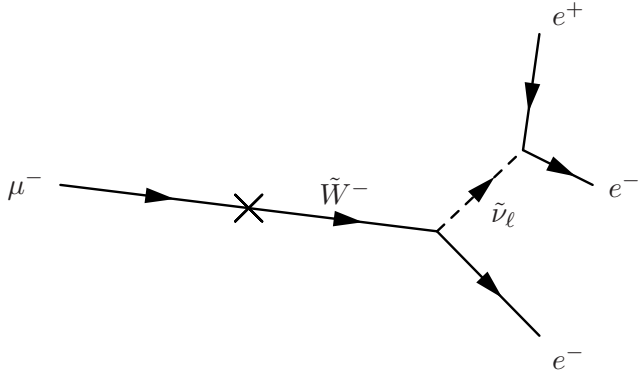
\begin{figure}
\begin{center}
\begin{fmfgraph*}(70,40)
\fmfleft{i}
\fmfright{o1,o2,o3}
\fmflabel{$\mu^{-}$}{i}
\fmflabel{$e^{-}$}{o1}
\fmflabel{$e^{-}$}{o2}
\fmflabel{$e^{+}$}{o3}
\fmf{fermion}{i,v1}
\fmf{fermion,label=$\tilde{W}^{-}$}{v1,v2}
\fmf{fermion}{v2,o1}
\fmf{scalar,label=$\tilde{\nu}_{\ell}$}{v2,v3}
\fmf{fermion}{v3,o2}
\fmf{fermion}{o3,v3}
\fmfv{decoration.shape=cross}{v1}
\fmfv{decoration.shape=dot}{v2,v3}
\end{fmfgraph*}
\end{center}
\caption{Muon decay to two electrons and a positron in SRPV NMSSM.\label{fig:muto3e}}
\end{figure}
\end{fmffile}

In the model considered here, the leading tree-level contribution to $\mu \to 3e$ is the process shown in Figure \ref{fig:muto3e}.
The low-energy ef\mbox{}fective superpotential looks like explicit bilinear RPV. The bilinear terms can be rotated away by def\mbox{}ining new combinations of lepton superf\mbox{}ields and $H_{d}$. We then have ef\mbox{}fective trilinear RPV terms. The ef\mbox{}fective $\lambda_{i11}$ will have a value $h_{e}v_{\nu i}/v_{d}$, where $h_{e}$ is the electron Yukawa coupling.
The lepton-wino mixing is essentially limited by neutrino masses. If we denote the mixing of the leptonic state by $\cos\gamma |\ell^{-}\rangle +\sin\gamma |\tilde{W}^{-}\rangle$, we find from our scan that the value of $\sin \gamma$ is less than $10^{-4}$ and typically somewhere around $10^{-6}$. The intermediate particle is a left-handed sneutrino. Its mass can be determined from the vacuum condition and it varies significantly in our data set.
Typical values are around $1$~TeV, but the lightest values are slightly below $100$~GeV. Taking the electrons massless and assuming this to be the dominant contribution we estimate the branching ratio in the limit of large $\tan \beta$ to be
\begin{equation}
BR(\mu\to eee)=3\cdot 10^{-30}\times \left(\frac{v_{\nu}}{1\;\mathrm{MeV}}\right)^{2}\left(\frac{\sin \gamma}{10^{-4}} \right)^{2} \left( \frac{100\; \mathrm{GeV}}{m_{\tilde{\nu}}} \right)^{4} \tan^{4} \beta.
\end{equation}
Although the branching ratio may be enhanced due to the $\tan \beta$ dependence, it will still remain several orders of magnitude below the experimental limits, even for the proposed improvement to 
$BR(\mu \to eee) \sim {\cal{O}}(10^{-16})$ \cite{Mu3e,Blondel:2013ia}. 
The leptonic f\mbox{}lavor constraints are avoided essentially because the reactions forbidden in the SM are mediated by left-handed sneutrinos, whose VEVs and thus also mixing is strongly constrained by the neutrino masses. The singlet sneutrinos, whose VEVs may be large, do not have gauge interactions and the Yukawa couplings are also small. 

Constraints from B-decays depend strongly on the Higgs sector spectrum.
The latest average for the experimental result of $b\rightarrow s\gamma$ is given by \cite{HFG}
$$\mathrm{BR}(b\to s\gamma) =343\pm 21\pm 7 \cdot 10^{-6}.$$ 
The branching ratio depends on the charged Higgs and chargino masses.
In the charged Higgs sector, charged doublets mix with charged sleptons, but the mixings 
are all suppressed by $h^{\nu}$, $v_{\nu}$ or $A_{\nu}$.
For all the scanned points, the charged Higgs mass will always be above $1$~TeV due to the large $A$-terms needed for spontaneous RPV. 
The constraints on the charged Higgs mass from f\mbox{}lavor physics, especially from $\mathrm{BR}(b\to s\gamma)$, require $m_{H^{\pm}}>316$~GeV \cite{Deschamps:2009rh}, which is clearly satisf\mbox{}ied.

The LHCb and CMS have measured the branching ratio of the rare process \cite{CMSandLHCbCollaborations:2013pla}
$$\mathrm{BR}(B_{s}\to \mu^{+}\mu^{-})=(2.9 \pm 0.7)\times 10^{-9}.$$
This is compatible with the Standard Model prediction $(3.56 \pm 0.18)\times 10^{-9}$ \cite{Buras:2013uqa}. 
In R-parity violating models there are several new sources for this reaction both at tree-level and at one-loop level and the limits on R-parity violating couplings have been analyzed \cite{Li:2013fa, Dreiner:2013jta}. 
The most stringent bounds on the products of two trilinear couplings are of the order of $3\cdot 10^{-11}\times m_{\tilde{\nu}}^{2}$, where the masses are given in GeV's. Assuming $m_{\tilde{\nu}}>100$~GeV we get bounds of the order of $10^{-4}$ for the trilinear R-parity violating couplings.



In spontaneous R-parity violation the ef\mbox{}fective trilinear R-parity violating couplings are Yukawa couplings multiplied by $v_{\nu}/v \cos \beta$. Since $v_{\nu}$ is very constrained by neutrino masses, the R-parity violating couplings are tiny. Taking the b-quark Yukawa coupling and $v_{\nu}=1$~MeV we get $\lambda'\simeq 10^{-7}\times \tan \beta$ in the limit of large $\tan \beta$. Hence the R-parity violating couplings satisfy the bounds from $B_{s} \to \mu\mu$ but on the other hand cannot produce a large deviation from the SM value, either.

\begin{figure}
\begin{center}
\includegraphics[width=\textwidth]{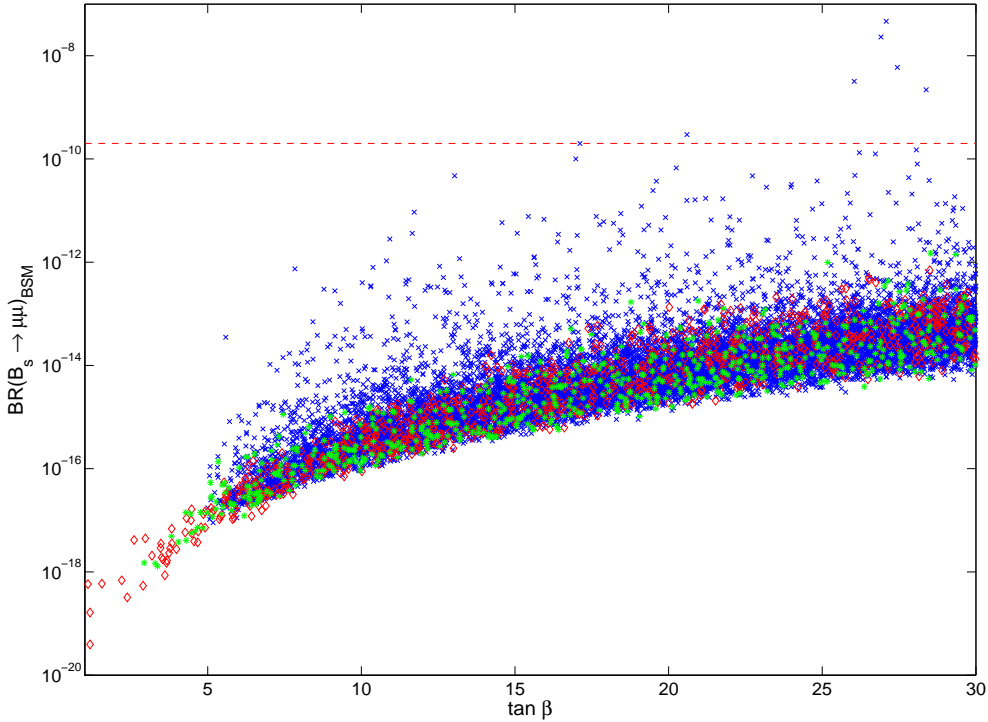}
\end{center}
\caption{The contribution from the CP-odd Higgs to BR$(B_{s}\to \mu\mu)$ as a function of $\tan \beta$. The dashed line gives the experimental $2\sigma$ limit assuming constructive interference with the SM amplitude.
The color coding is same as in Figure \ref{fig:sneutmix}.\label{fig:bsmumu}}
\end{figure}

The leading contribution to the reaction $B_{s}\to \mu\mu$ is by the neutral Higgs bosons \cite{Babu:1999hn}. Especially the contribution from the CP-odd doublet Higgs may be large. We estimated the contribution of the CP-odd Higgs conservatively by choosing the state which is mostly the imaginary part of $H_{d}$, assuming that it has the same couplings as in the MSSM, used the large $m_{A}$, $\tan \beta$ limit (see \cite{Babu:1999hn}) and assumed constructive interference with the SM amplitude. 
In Figure \ref{fig:bsmumu} we have shown the contribution from the CP-odd Higgs. 
When the contributions of this and the SM are summed at the amplitude level and constructive interference is assumed, the contribution to the branching ratio from the CP-odd Higgs must be below $2\times 10^{-10}$ for the total branching ratio to be below $4.3\times 10^{-9}$, i.e. $2\sigma$ over the central value. 
In our data set, when the CP-odd Higgs is doublet dominated, its mass
is typically a few TeV's which suppresses the branching ratio to unobservable values.
We may note that although typically the experimental constraints are satisf\mbox{}ied, this contribution may lead to larger deviations from the SM prediction than the ef\mbox{}fective trilinear R-parity violating couplings.
When the CP-odd doublet Higgs is light and $\tan \beta$ is large we have a few data points, where this contribution would be experimentally ruled out.

\subsection{Higgs couplings and decays}
\label{sec:couplings}

The most important tree-level couplings of the SM-like Higgs are to the heavy third generation quarks and to the W/Z bosons. The former are essential both in the production of Higgs through gluon-gluon fusion and in the decay to photons, while the
couplings to gauge bosons are especially important in the decays, both to photons via a W-loop, and 
$h\rightarrow ZZ\rightarrow$ charged leptons, which gives the most straightforward means to measure the Higgs mass.

\begin{figure}
\begin{center}
\includegraphics[width=0.87\textwidth]{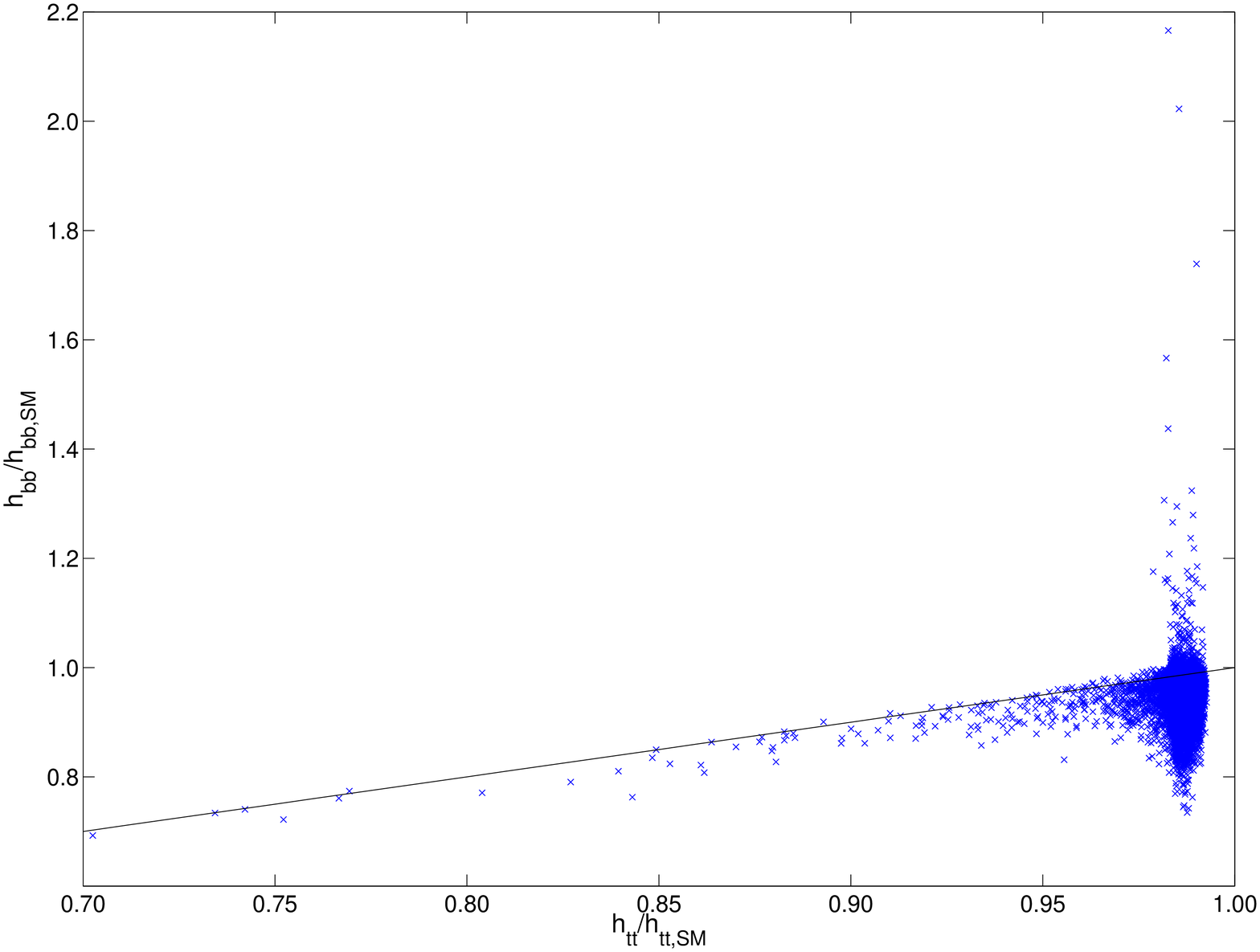}
\includegraphics[width=0.87\textwidth]{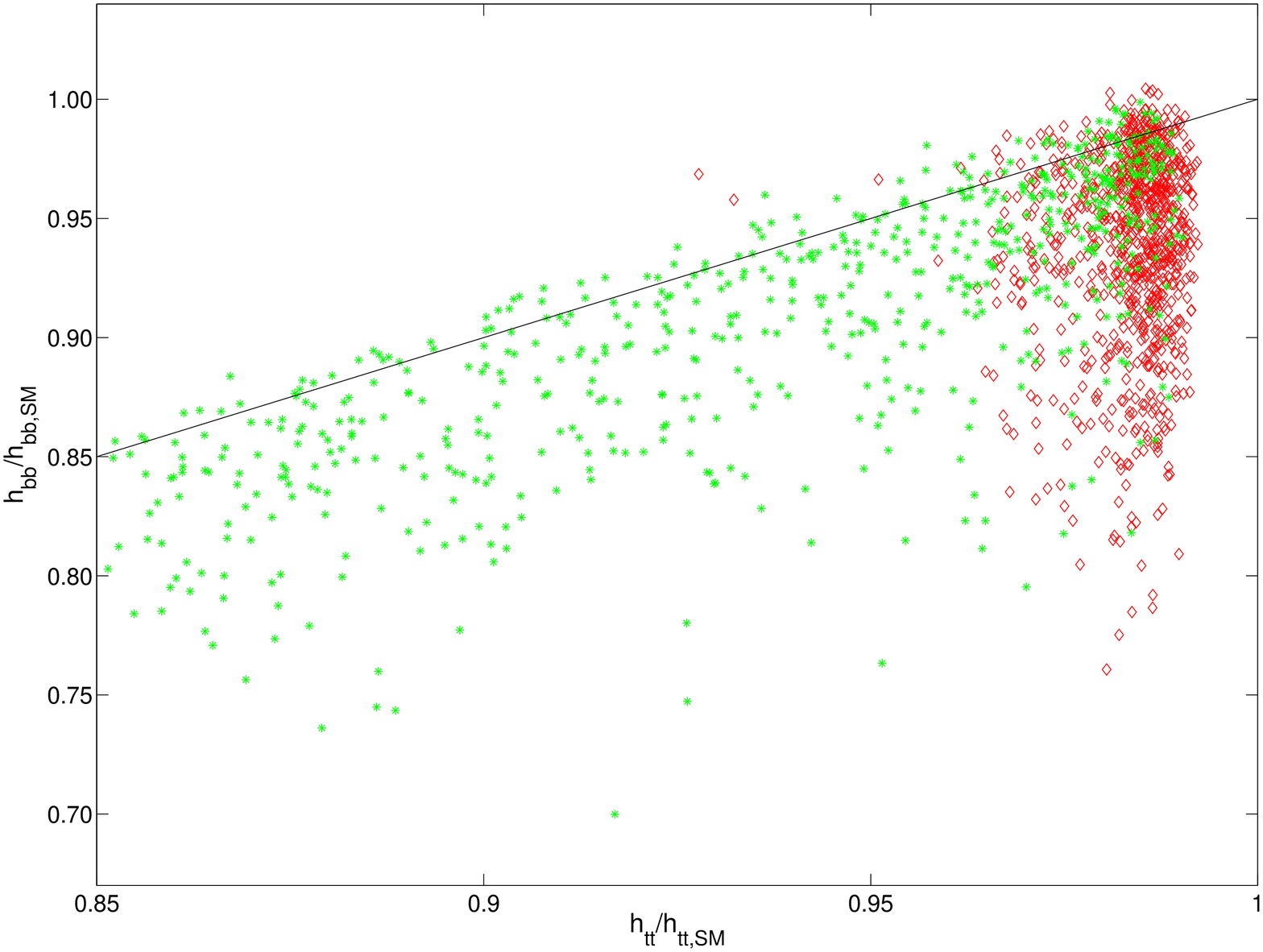}
\end{center}
\caption{The Higgs-bottom coupling ratio to the SM value as a function of the corresponding Higgs-top coupling ratio if the SM-like Higgs is the lightest scalar (upper) or second lightest (lower). The black line shows where the suppression of couplings is equal. 
The color coding is same as in Figure \ref{fig:sneutmix}.
\label{fig:quarkc1}}
\end{figure}

The couplings to the third generation quarks compared to the SM values are plotted in Figure~\ref{fig:quarkc1}.
In the quark couplings we added leading one-loop SUSY-QCD corrections \cite{Hall:1993gn, Hempfling:1993kv,Carena:1994bv,Pierce:1996zz} with $M_{\tilde{g}}=1.5$~TeV. 
Since the lightest pseudoscalar for majority of the points is heavy, due to the decoupling effect the tree level couplings are mostly close to the Standard Model values.
However, the b-quark couplings with large $\tan\beta$ can have sizable SUSY-QCD corrections, while corrections for the top quark couplings are not large \cite{Han:2013sga}.
Thus, even if the coupling to bottoms is strongly enhanced, the coupling to the top-quarks is SM like.\footnote{This
has been studied in the MSSM, when the CP-even Higgses are nearly degenerate in mass \cite{Boos:2002ze}.
In such a case the coupling to bottoms can be increased by a large amount.} 
Universal suppression of the couplings due to the doublet-sneutrino mixing is clearly visible for both b- and t-quark couplings.

If there is a lighter scalar than the $125$~GeV particle the b-quark coupling cannot be enhanced  and a $10\%$ suppression (due to SUSY-QCD corrections) is typical. This behaviour can be understood as follows.
Consider again the linear combinations $h=\sin \beta H_{u}^{0}+\cos \beta H_{d}^{0}$ and $H=\cos \beta H_{u}^{0}-\sin \beta H_{d}^{0}$. The f\mbox{}irst one has the same VEV and couplings as the SM Higgs  and the second one has zero VEV. The enhancement of the b-quark coupling is due to the mixing of $h$ and $H$.
Look then at the mass matrix of Eq. (\ref{eq:hHNmatrix}) and start by diagonalizing the submatrix formed by $H$ and $\tilde{N}$. Since the mass splitting of the eigenstates is always larger than the dif\mbox{}ference of the diagonal elements, the one with the smaller diagonal element becomes lighter and the one with the larger diagonal element becomes heavier. If there is a light sneutrino-dominated state as in the lower plot of Figure \ref{fig:quarkc1}, the other doublet state becomes heavier than the diagonal element and so it decouples and cannot mix substantially with the SM-like Higgs. Thus there cannot be a large enhancement in the b-quark coupling if we have a light sneutrino-like state.
If the sneutrino-dominated state is heavy, i.e. $\lambda_{N}$ and $v_{N}$ are large, the mixing between $H$ and $\tilde{N}$ forces the $H$-dominated state to be relatively light. Hence it can mix with the SM-like Higgs state and a large enhancement in the b-quark coupling is possible.

A similar ef\mbox{}fect happens also in the CP-odd mass matrix. The light doublet dominated states at large values of $\lambda_{N}$ in Figure \ref{fig:ma} are due to a heavy CP-odd sneutrino and the mixing term between the sneutrino and the doublet. In this case the light CP-odd doublet Higgs may also enhance the branching ratio of $B_{s}\to \mu\mu$. Hence we f\mbox{}ind a correlation between a large deviation of BR$(B_{s}\to\mu\mu)$ from its SM value and a large enhancement in the coupling between the Higgs and the b-quarks. Since in the charged Higgs mass matrix such a mixing cannot happen, the charged Higgs may be significantly heavier than the MSSM-type CP-odd and heavy CP-even Higgses.

\begin{figure}
\begin{center}
\includegraphics[width=\textwidth]{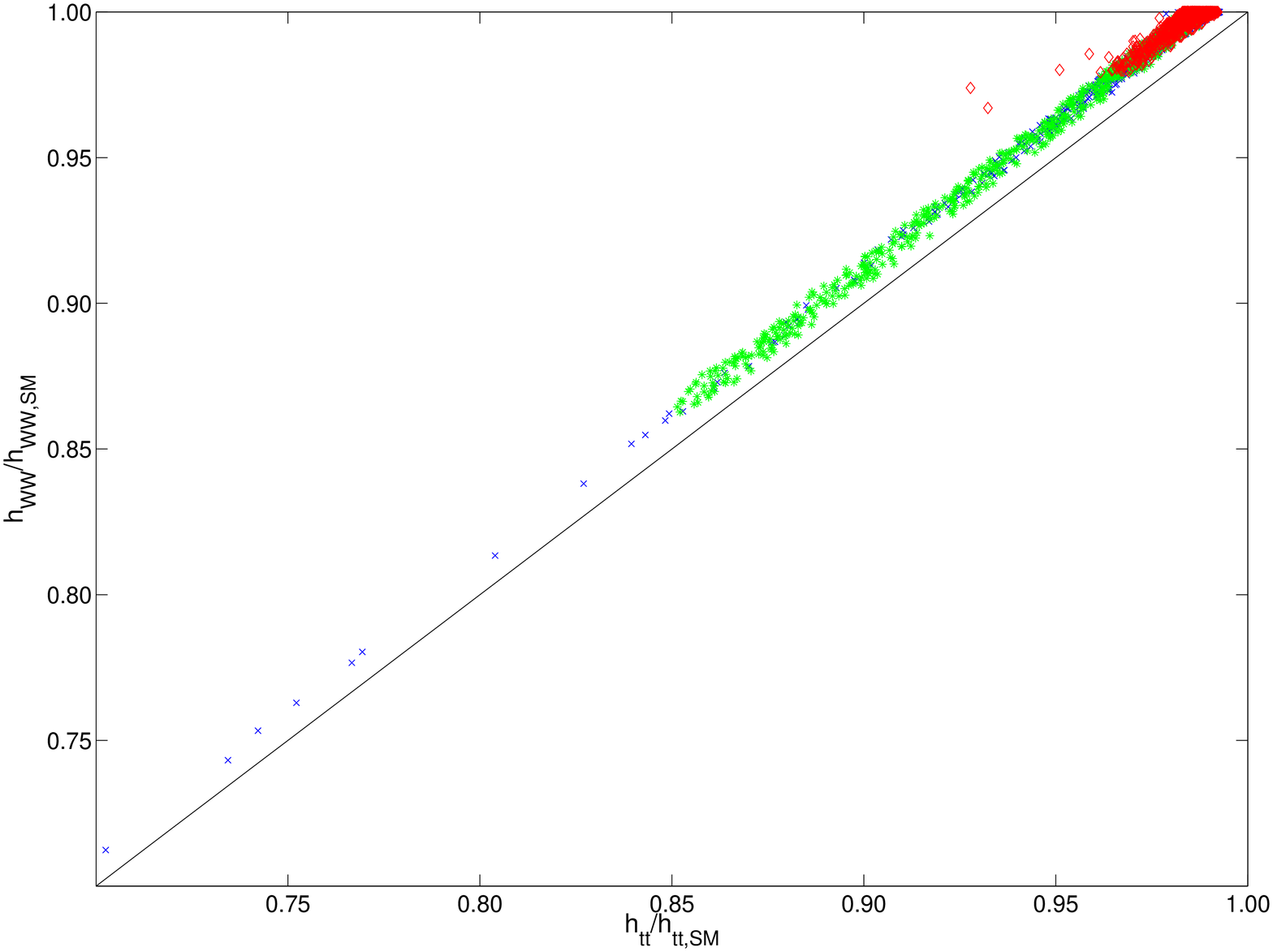}
\end{center}
\caption{The ratio of the coupling to W/Z compared to the SM as a function of the same ratio of top couplings. The color coding is same as in Figure \ref{fig:sneutmix}.\label{fig:WWcoup}}
\end{figure}

In Fig.~\ref{fig:WWcoup} a similar ratio for W-bosons and top quarks is shown. 
We see that the couplings are strongly correlated. 
Deviations from the SM values are mostly due to the doublet-sneutrino mixing
because the relevant parameter space is close to the decoupling limit as
discussed earlier.
All the points are slightly above the diagonal due to the SUSY-QCD corrections
to the top Yukawa couplings.

The most discussed deviation at the LHC from the SM Higgs prediction has been the possible higher rate in the two-photon decay channel \cite{ATLAS:2012ad, Chatrchyan:2012tw, Ellwanger:2011aa, Ferreira:2012my, Baglio:2012et, Giudice:2012pf}. The SM contribution to this channel is dominated by the W-boson loop and with a subleading destructive contribution from the top quark loop. 

As we have seen, in the SRPV-NMSSM model the SM-like Higgs may have a substantial sneutrino component.
The decay mode of $\tilde N$ to two photons via a charged Higgs loop is not suppressed by mixing in the Higgs sector, or by tiny parameters.
This contribution can be either constructive or destructive depending on the relative signs of the components of the corresponding eigenvector.
However, in general, any scalar loop will give only a minor contribution to the amplitude compared to vector loops. As discussed earlier, the charged Higgs is heavier than 1 TeV for our data set so the loop will be suppressed for that reason also. The modification from the sneutrino component on the branching ratio BR$(h\rightarrow \gamma\gamma)$ will be less than $1\%$. With the current experimental precision this is indistinguishable.

The important contributions that can alter the two-photon rate, are the coupling to the top quark, which determines the gluon fusion production rate and the coupling to the bottom quark, which gives the main component of the total decay width. If the ratio of the top coupling to its SM value is larger than that of the bottom coupling, the two-photon rate may be enhanced. This corresponds to points which are below the black lines in Figure \ref{fig:quarkc1}. 
The top coupling is typically very close to the SM value (or there is a universal suppression due to the sneutrino component) but the bottom coupling has a larger deviation due to SUSY-QCD corrections proportional to the top Yukawa coupling.

\begin{figure}
\begin{center}
\includegraphics[width=0.9\textwidth]{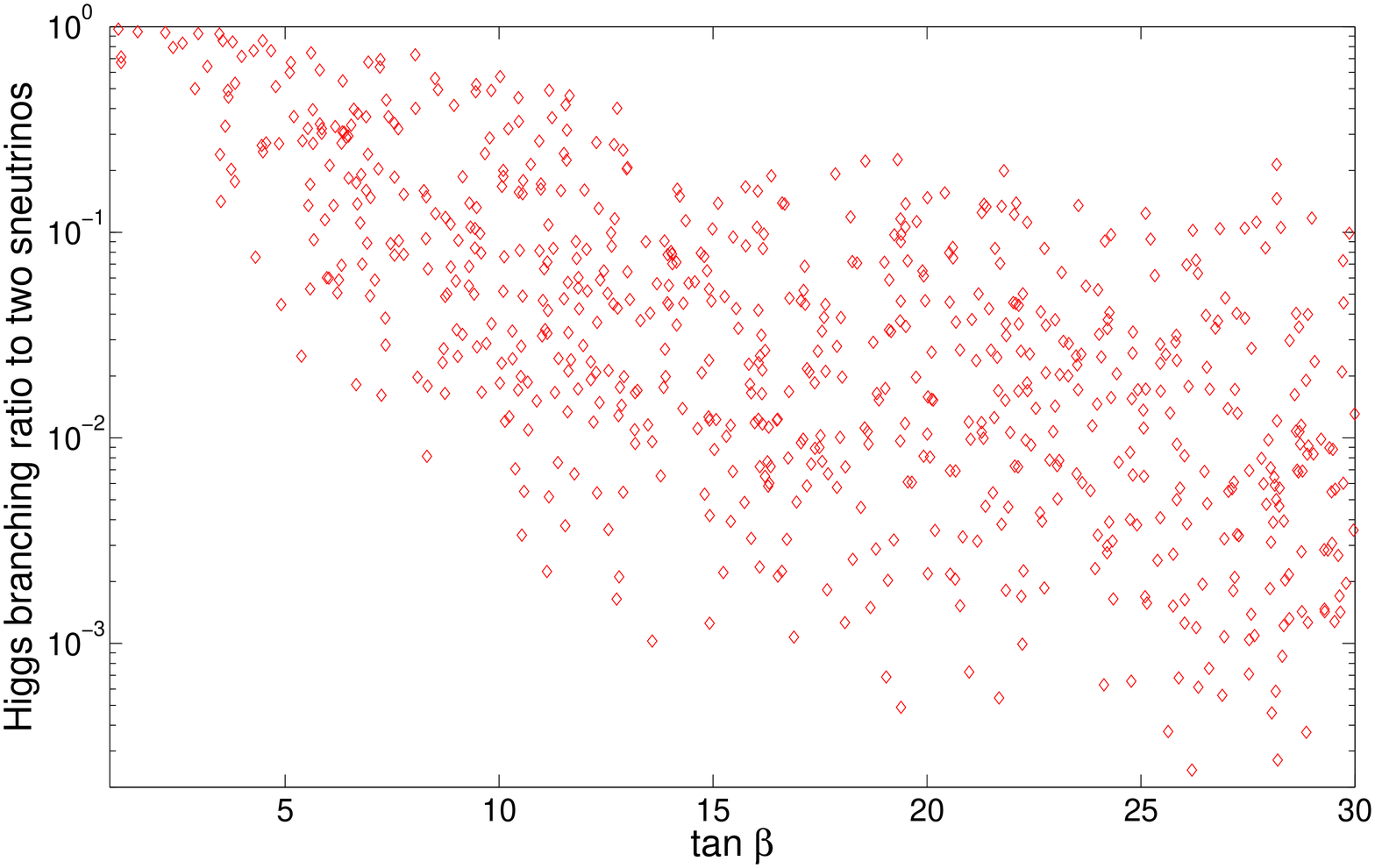}
\end{center}
\caption{The Higgs branching ratio to two sneutrinos as a function of $\tan \beta$ for the data points with a sneutrino-dominated state lighter than $60$~GeV.\label{fig:sneutrinowidth}}
\end{figure}

If the sneutrino is lighter than $m_{h}/2$, the total decay width is af\mbox{}fected by the decay of the SM-like Higgs to two sneutrinos. The coupling between the SM Higgs and a pure sneutrino is $\frac{1}{4}\lambda_{H}\lambda_{N}v\sin 2\beta$ and is thus suppressed at high values of $\tan \beta$ but large in the limit $\tan \beta \rightarrow 1$. The branching ratio $BR(h\rightarrow \tilde{N}\tilde{N})$, shown in Figure \ref{fig:sneutrinowidth}, depends on $m_{\tilde{N}}$, $\lambda_{N}$, $\lambda_{H}$ and $\tan \beta$ and it can vary over several orders of magnitude and even be the dominant decay mode. This would lead to a large suppression of all other decay modes. At large values of $\tan \beta$ the branching ratio can be less than $10^{-3}$. Since the Higgs signals are not largely suppressed, we can exclude the region $\tan \beta \lesssim 3$ in the case of a light sneutrino.

\begin{figure}
\begin{center}
\includegraphics[width=0.87\textwidth]{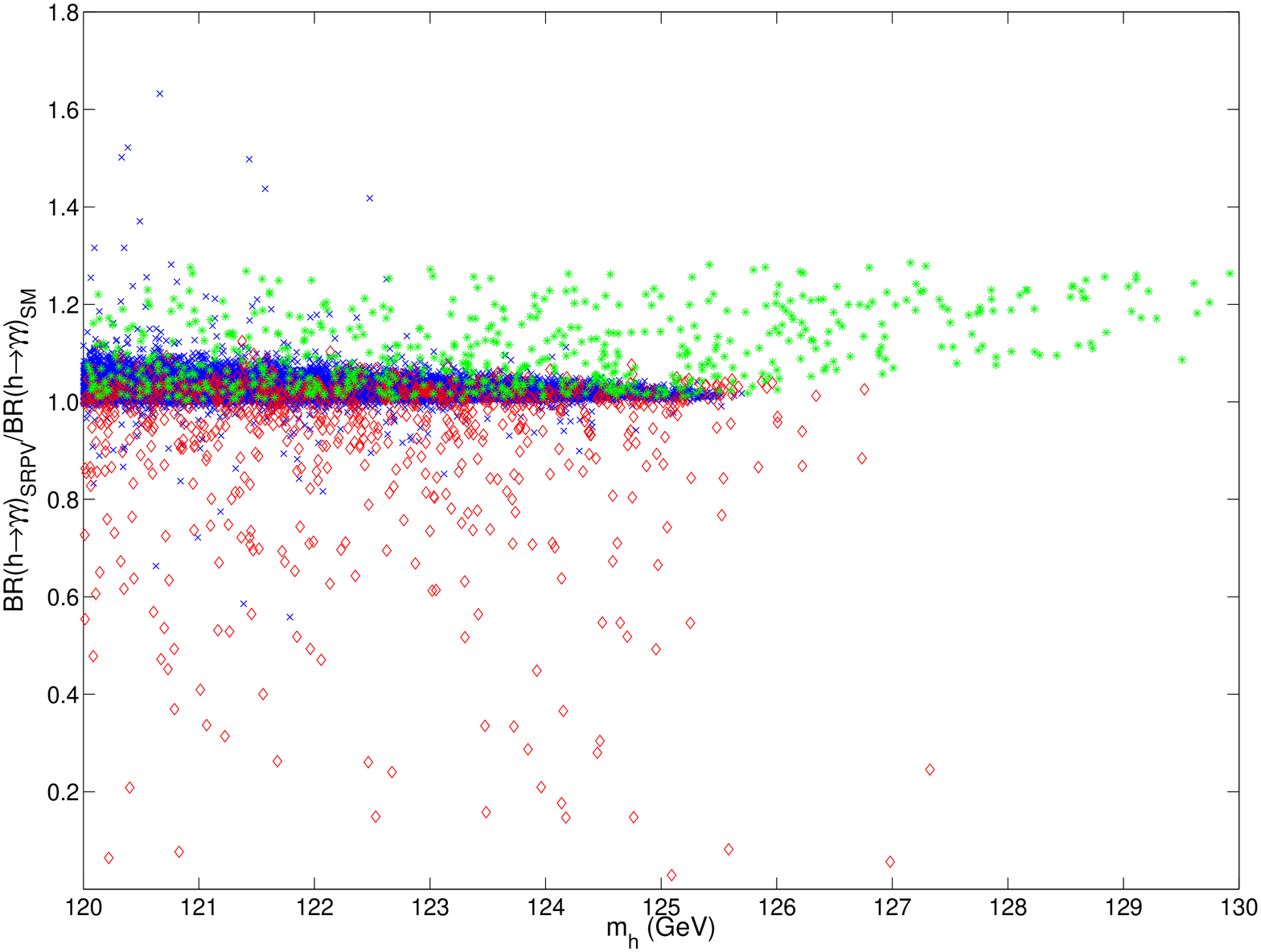}
\includegraphics[width=0.87\textwidth]{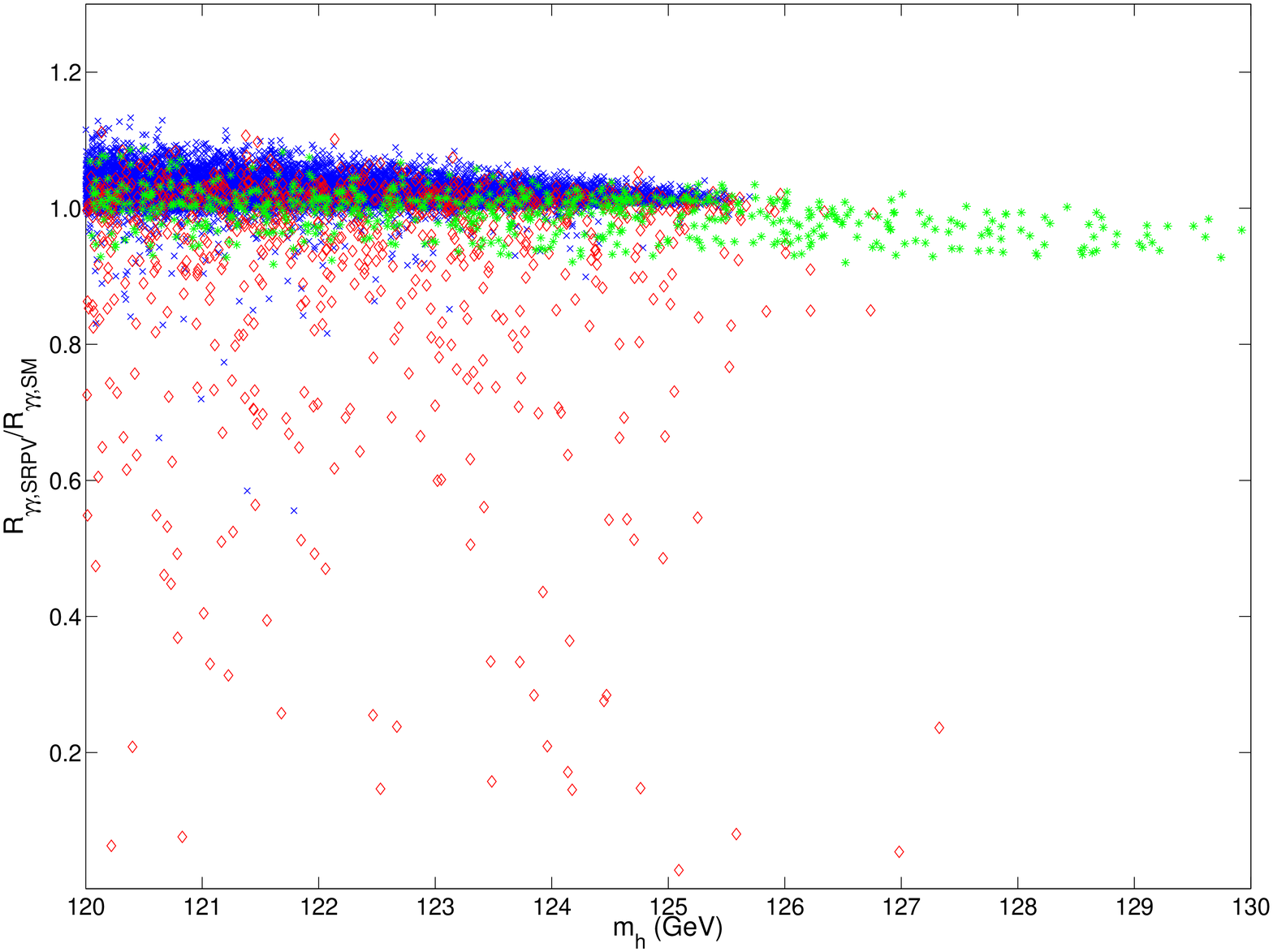}
\end{center}
\caption{The branching ratio (upper) and the event rate (lower) to the two-photon f\mbox{}inal state compared to the SM prediction as a function of the SM-like Higgs mass. An enhancement in the branching ratio is possible in this model but the event rate is close to the SM prediction. These plots are computed for production by vector boson fusion (or associated production). The color coding is same as in Figure \ref{fig:sneutmix}.\label{fig:BRgamgam}}
\end{figure}

The branching ratio of the two-photon channel compared to the SM-value is shown in the upper plot of Figure \ref{fig:BRgamgam}. 
We see that an enhancement is possible especially if there is a lighter scalar between $80$ and $120$~GeV. 
However the gluon fusion or vector boson fusion cross section is suppressed because of the doublet-sneutrino mixing.
Hence the overall rates are close to the Standard Model expectation as is shown in the lower plot of Figure \ref{fig:BRgamgam}.

Another possible reason for an increased two-photon rate could be that the sneutrino dominated scalar and the SM-like Higgs are almost degenerate in mass. 
Such a phenomenon has been studied in the context of RPC NMSSM, see \cite{Gunion:2012gc}.
The sneutrino state will have a large branching fraction to two photons since there are not many f\mbox{}inal states that would have a large coupling and be kinematically allowed. However, the sneutrino is not produced at a rate comparable to the SM-like Higgs.
The sneutrino has unsuppressed couplings only to the singlet Higgs, doublet Higgses and the charged Higgs. 
For a mixed sneutrino-Higgs state the production will be dominated by the Higgs component even though it were small.

In the case of a heavy sneutrino, the sneutrino-dominated state can be produced if it mixes with the doublet Higgses, since it has unsuppressed couplings only to the singlet and doublet Higgses. If the sneutrino is heavy enough, the decay $\tilde{N}\rightarrow hh$ is allowed on-shell.
If all of the Higgses are heavier than the sneutrino, it will decay mostly to a pair of off-shell Higgs bosons, which then subsequently decay. If the sneutrino state mixes with doublet Higgses the production will be determined by the doublet Higgs component but the sneutrino decay modes may be competitive to the doublet Higgs component decay modes. 

If the sneutrino is very light, say a few GeV's, the most usual decay modes through the Higgs bosons become kinematically forbidden. In that case the sneutrino could decay to two almost collinear photons, which could be seen as one in the detector. This could lead to a change in the observed two-photon rate also.

In order to better understand the consistency of our results with the current experimental results, we study two subsets of our data, where the Higgs mass is enhanced.
In one subset we require $m_{\tilde{t}}< 900$~GeV and in the other $m_{h}>126$~GeV, both having a sneutrino dominated state between 80 and 120~GeV. The former one has 50 data points and the latter 112 data points. 
The diphoton rate in these samples is between 0.92 and 1.09 times the Standard Model value. The combined early results of ATLAS and CMS, which are taken into the PDG average \cite{Agashe:2014kda} of $1.58^{+0.27}_{-0.23}$ times the Standard Model value leave this dataset outside the $95\%$ confidence level. The latest ATLAS and CMS results \cite{Aad:2014eha, Khachatryan:2014ira}, not included in the PDG average, give signal strengths $1.17\pm 0.27$ and $1.14^{+0.26}_{-0.23}$, respectively. Our datasets are within the $68\%$ conf\mbox{}idence level of these results.
In these datasets the top-quark coupling is $0.85\ldots 0.98$, the bottom-quark coupling is $0.76\ldots 0.97$ and the coupling to vector bosons $0.86\ldots 0.98$ times the Standard Model value. The Higgs-sneutrino mixing leads to a correlation in the suppressions. After the Higgs discovery a few model-independent fits on the Higgs couplings have been done \cite{Djouadi:2013qya,Belanger:2013xza}. Our results mostly are within the $68\%$ conf\mbox{}idence level of these fits. This is not surprising since the fermionic couplings still have large uncertainties but they will be reduced in the next run of the LHC.

\section{Conclusions}
\label{sec:conclusions}

It is of interest to study consequences of the dynamical breaking of R-parity,
contrary to the explicit breaking, which is the only possibility in MSSM.
MSSM is under pressure because of the fine-tuning issues, since no supersymmetric
partners have been found at the LHC.
In this work we have studied the scalar sector of a spontaneously R-parity violating model, 
which is NMSSM extended by an explicitly lepton number
violating term and a term due to nonrenormalizable terms.
NMSSM is the simplest extension of the minimal supersymmetric standard model.
Although spontaneously broken R-parity is not possible in the MSSM, it can occur
in the NMSSM with right-handed neutrinos and a tadpole term which takes care of the cosmological domain wall problem.

We have concentrated on the properties of the SM-like Higgs
when relevant constraints have been taken into account.
We have found that the lightest scalar in the model may be a sneutrino, which
helps to make the SM-like Higgs heavier that in the MSSM.
Thus the stop can be lighter than typically in MSSM and could be discovered
in the 14 TeV phase of the LHC.
Interestingly, in SRPV-NMSSM, the $A$-terms are always large, since
otherwise the vacuum would be R-parity conserving.

The identity of the light scalars in various NMSSM-models differ. 
It is possible to have a lighter than 125 GeV scalar also in the R-parity conserving
NMSSM, but in that case it is dominantly the other singlet of the model.
In both cases it is also possible not to have an additional light scalar.

We found that in the case of a light sneutrino-like state the Higgs coupling to b-quarks cannot be enhanced compared to the SM. If the sneutrino is heavy, the mixing between the sneutrino and the doublet Higgses may make both of the CP-even doublet Higgses light and their mixing can enhance the b-quark coupling. In this case also the CP-odd doublet state becomes light, wheras the charged Higgs is signif\mbox{}icantly heavier.

The Higgs decay rate to a photon pair may be enhanced or suppressed but
in our data set the differences to the SM Higgs are not large.
The best possibilities to identify the model may be through a light
sneutrino dominated scalar, if the doublet component is large enough
for it to be produced in significant amounts.
We will leave this study for a future work.

We found that in the case where the Higgs-sneutrino mixing lifts the SM-like Higgs mass the Higgs couplings are within the current uncertainties of model-independent Higgs coupling fits. The next LHC run should reduce these uncertainties to a level where the suppression of Higgs couplings due to a sneutrino component could be seen.

\section*{Acknowledgements}

The authors acknowledge support from the Academy of Finland (Project Nro 137960).
We thank Gautam Bhattacharyya for several useful discussions concerning stop.


\section*{Appendix}
\appendix
\section{Minimization conditions and CP-odd scalar mass matrix}\label{AppA}

Minimizing the tree-level potential gives the following conditions:
\begin{eqnarray}
\frac{\partial V}{\partial H_{u}^{0}} & = & \frac{1}{2}A_{H}v_{d}v_{\Phi}+\frac{v_{u}}{\sqrt{2}}\left( 2m_{H_{u}}^{2}+\lambda_{H}^{2}(v_{\Phi}^{2}+v_{d}^{2})-m_{Z}^{2}\cos 2\beta \right)\nonumber \\
& & +\frac{v_{d}}{\sqrt{2}}\left( \frac{1}{2}\lambda_{H}\lambda_{N}v_{N}^{2}+\frac{1}{2}\lambda_{H}\lambda_{\Phi}v_{\Phi}^{2} \right)=0\label{eq:humin},\\
\frac{\partial V}{\partial H_{d}^{0}} & = & \frac{1}{2}A_{H}v_{u}v_{\Phi}+\frac{v_{d}}{\sqrt{2}}\left( 2m_{H_{d}}^{2}+\lambda_{H}^{2}(v_{\Phi}^{2}+v_{u}^{2})+m_{Z}^{2}\cos 2\beta \right)\nonumber \\
& & +\frac{v_{u}}{\sqrt{2}}\left( \frac{1}{2}\lambda_{H}\lambda_{N}v_{N}^{2}+\frac{1}{2}\lambda_{H}\lambda_{\Phi}v_{\Phi}^{2} \right)=0,\\
\frac{\partial V}{\partial \Phi} & = & 2\xi^{3}+\frac{1}{2}A_{N}v_{N}^{2}+A_{H}v_{u}v_{d}+\frac{1}{2}A_{\Phi}v_{\Phi}^{2}+\frac{v_{\Phi}}{\sqrt{2}} (2m_{\Phi}^{2}\nonumber \\
& & +\lambda_{H}^{2}v^{2}+\lambda_{N}^{2}v_{N}^{2} +\lambda_{\Phi}\lambda_{H}v_{d}v_{u}+\frac{1}{2}\lambda_{\Phi}\lambda_{N}v_{N}^{2}+\frac{1}{2}\lambda_{\Phi}^{2}v_{\Phi}^{2})=0,\\
\frac{\partial V}{\partial \tilde{N}} & = & \frac{v_{N}}{\sqrt{2}}( 2m_{N}^{2}+2A_{N}\frac{v_{\Phi}}{\sqrt{2}}+\lambda_{N}^{2}v_{\Phi}^{2}+\lambda_{N}\lambda_{H}v_{d}v_{u} \nonumber\\
& & +\frac{1}{2}\lambda_{N}^{2}v_{N}^{2}+\frac{1}{2}\lambda_{N}\lambda_{\Phi}v_{\Phi}^{2})=0,\label{eq:sneutrinomin}\\
\frac{\partial V}{\partial \tilde{\nu}} & = & \frac{v_{\nu}}{\sqrt{2}}\left( 2m_{\tilde{L}}^{2}+m_{Z}^{2}\cos 2\beta \right)+h^{\nu}v_{\Phi}v_{N}(\lambda_{N}v_{u}+\lambda_{H}v_{d})/\sqrt{2}\nonumber\\
& & +A_{\nu}v_{u}v_{N} = 0\label{eq:lsneutrinomin}.
\end{eqnarray}

We have assumed that $h^{\nu}$, $v_{\nu}$ and $A_{\nu}$ are small and terms containing may be neglected in sums which have also nonsuppressed terms. We use these relations to eliminate the soft scalar masses. This leads to the following matrix elements for the tree-level CP-odd scalar mass-squared matrix
\begin{eqnarray}
m_{11}^{2} & = & -\frac{1}{2}(A_{H}v_{\Phi}/\sqrt{2}+\lambda^{2}V^{2})\cot \beta,\\
m_{22}^{2} & = & -\frac{1}{2}(A_{H}v_{\Phi}/\sqrt{2}+\lambda^{2}V^{2})\tan \beta,\\
m_{33}^{2} & = & -\frac{\overline{\xi}^{3}}{v_{\Phi}/\sqrt{2}}-\frac{3}{2}A_{\Phi}\frac{v_{\Phi}}{\sqrt{2}}-\frac{1}{2}\lambda_{\Phi}\lambda_{N}v_{N}^{2} -\lambda_{H}\lambda_{\Phi}v_{u}v_{d},\\
m_{44}^{2} & = & -2A_{N}\frac{v_{\Phi}}{\sqrt{2}}-\frac{1}{2}\lambda_{\Phi}\lambda_{N}v_{\Phi}^{2}-\lambda_{H}\lambda_{N}v_{u}v_{d},\label{eq:cpodd44}\\
m_{12}^{2} & = & -\frac{1}{2}(A_{H}v_{\Phi}/\sqrt{2}+\lambda^{2}V^{2}),\\
m_{13}^{2} & = & -\frac{1}{2}A_{H}\frac{v_{d}}{\sqrt{2}}+\frac{1}{2}\lambda_{H}\lambda_{\Phi}v_{d}v_{\Phi},\\
m_{14}^{2} & = & \frac{1}{2}\lambda_{H}\lambda_{N}v_{d}v_{N},\\
m_{23}^{2} & = & -\frac{1}{2}A_{H}\frac{v_{u}}{\sqrt{2}}+\frac{1}{2}\lambda_{H}\lambda_{\Phi}v_{u}v_{\Phi},\\
m_{24}^{2} & = & \frac{1}{2}\lambda_{H}\lambda_{N}v_{u}v_{N},\\
m_{34}^{2} & = & -\frac{1}{2}A_{N}\frac{v_{N}}{\sqrt{2}}+\frac{1}{2}\lambda_{N}\lambda_{\Phi}v_{N}v_{\Phi}.
\end{eqnarray}
The notations $v^2=v_{u}^{2}+v_{d}^{2}=(246\;\mathrm{GeV})^{2}$, $\lambda^{2}V^{2}=\frac{1}{2}(\lambda_{H}\lambda_{N}v_{N}^{2}+\lambda_{H}\lambda_{\Phi}v_{\Phi}^{2})$ and $\overline{\xi}^{3}=\xi^{3}+\frac{1}{4}A_{N}v_{N}^{2}+\frac{1}{2}A_{H}v_{u}v_{d}$ have been used.
All of the parameters have been assumed real for simplicity.


\begin{thebibliography}{99}

\bibitem{Chatrchyan:2012ufa}
  S.~Chatrchyan {\it et al.}  [CMS Collaboration],
  Phys.\ Lett.\ B {\bf 716} (2012) 30
  [arXiv:1207.7235 [hep-ex]].
  
\bibitem{Aad:2012tfa}
  G.~Aad {\it et al.}  [ATLAS Collaboration],
  Phys.\ Lett.\ B {\bf 716} (2012) 1
  [arXiv:1207.7214 [hep-ex]].
  
\bibitem{ATLAS:2012ad}
  G.~Aad {\it et al.}  (ATLAS Collaboration),
  Phys.\ Rev.\ Lett.\  {\bf 108} (2012) 111803,
  arXiv:1202.1414 [hep-ex].

\bibitem{Chatrchyan:2012tw}
  S.~Chatrchyan {\it et al.}  (CMS Collaboration),
  Phys.\ Lett.\ B {\bf 710} (2012) 403,
  arXiv:1202.1487 [hep-ex].


\bibitem{Aad:2014eha}
  G.~Aad {\it et al.}  [ATLAS Collaboration],
  arXiv:1408.7084 [hep-ex].


\bibitem{Khachatryan:2014ira}
  V.~Khachatryan {\it et al.}  [CMS Collaboration],
  arXiv:1407.0558 [hep-ex].

\bibitem{Salam:1974xa}
  A.~Salam and J.~A.~Strathdee,
  Nucl.\ Phys.\ B {\bf 87} (1975) 85.
  
\bibitem{Fayet:1974pd}
  P.~Fayet,
  Nucl.\ Phys.\ B {\bf 90} (1975) 104.

\bibitem{Farrar:1978xj}
  G.~R.~Farrar and P.~Fayet,
  Phys.\ Lett.\ B {\bf 76} (1978) 575.
  
\bibitem{Hall:1983id}
  L.~J.~Hall and M.~Suzuki,
  Nucl.\ Phys.\ B {\bf 231} (1984) 419.

\bibitem{Barbier:2004ez}
  R.~Barbier, C.~Berat, M.~Besancon, M.~Chemtob, A.~Deandrea, E.~Dudas, P.~Fayet and S.~Lavignac {\it et al.},
  Phys.\ Rept.\  {\bf 420} (2005) 1
  [hep-ph/0406039].

 
\bibitem{Aulakh:1982yn}
  C.~S.~Aulakh and R.~N.~Mohapatra,
  Phys.\ Lett.\ B {\bf 119} (1982) 136.

\bibitem{Hayashi:1984rd}
  M.~J.~Hayashi and A.~Murayama,
  Phys.\ Lett.\ B {\bf 153} (1985) 251.

\bibitem{Mohapatra:1986aw}
  R.~N.~Mohapatra,
  Phys.\ Rev.\ Lett.\  {\bf 56} (1986) 561.

\bibitem{Masiero:1990uj}
  A.~Masiero and J.~W.~F.~Valle,
  Phys.\ Lett.\ B {\bf 251} (1990) 273.

\bibitem{Frank:2001tr}
  M.~Frank and K.~Huitu,
  Phys.\ Rev.\ D {\bf 64} (2001) 095015
  [hep-ph/0106004].

\bibitem{Grossman:1998py}
  Y.~Grossman and H.~E.~Haber,
  Phys.\ Rev.\ D {\bf 59} (1999) 093008
  [hep-ph/9810536].

\bibitem{Hirsch:2000ef}
  M.~Hirsch, M.~A.~Diaz, W.~Porod, J.~C.~Romao and J.~W.~F.~Valle,
  Phys.\ Rev.\ D {\bf 62} (2000) 113008
   [Erratum-ibid.\ D {\bf 65} (2002) 119901]
  [hep-ph/0004115].

\bibitem{Davidson:2000uc}
  S.~Davidson and M.~Losada,
  JHEP {\bf 0005} (2000) 021
  [hep-ph/0005080].

\bibitem{Abada:2002ju}
  A.~Abada, G.~Bhattacharyya and M.~Losada,
  Phys.\ Rev.\ D {\bf 66} (2002) 071701
  [hep-ph/0208009].

\bibitem{Mohapatra:1979ia}
  R.~N.~Mohapatra and G.~Senjanovic,
  Phys.\ Rev.\ Lett.\  {\bf 44}, 912 (1980);
  R.~N.~Mohapatra and G.~Senjanovic,
  Phys.\ Rev.\ D {\bf 23}, 165 (1981); 
M. Gell-Mann, P. Ramond, R. Slansky, Supergravity (P. van Nieuwenhuizen
et al. eds.), North Holland, Amsterdam, 1980, p. 315; 
T. Yanagida, in Proceedings of the Workshop on the Unified
Theory and the Baryon Number in the Universe (O. Sawada and
A. Sugamoto, eds.), KEK, Tsukuba, Japan, 1979, p. 95. 
S.L. Glashow, The future of elementary particle physics, in
Proceedings of the Summer Institute on Quarks and Leptons (M. Levy
et al eds.), Plenum Press, New York, 1980, pp. 687.

\bibitem{Valle}
J. Schechter, J.W.F. Valle, Phys.\ Rev.\ D {\bf 22}, 2227 (1980);
J. Schechter, J.W.F. Valle, Phys.\ Rev.\ D {\bf 25}, 774 (1982);
R.N. Mohapatra, J.W.F. Valle, Phys.\ Rev.\ D {\bf 34}, 774 (1986).

\bibitem{Kitano:1999qb}
  R.~Kitano and K.~-y.~Oda,
  Phys.\ Rev.\ D {\bf 61} (2000) 113001,
  hep-ph/9911327.


\bibitem{Frank:2005tn}
  M.~Frank, K.~Huitu and T.~Ruppell,
  hep-ph/0508056.

\bibitem{Frank:2007un}
  M.~Frank, K.~Huitu and T.~Ruppell,
  Eur.\ Phys.\ J.\ C {\bf 52} (2007) 413,
  arXiv:0705.4160 [hep-ph].

\bibitem{Cao:2011re}
  J.~-J.~Cao, K.~-i.~Hikasa, W.~Wang, J.~M.~Yang, K.~-i.~Hikasa, W.~-Y.~Wang and J.~M.~Yang,
  Phys.\ Lett.\ B {\bf 703} (2011) 292
  [arXiv:1104.1754 [hep-ph]].

\bibitem{Vasquez:2012hn}
  D.~A.~Vasquez, G.~Belanger, C.~Boehm, J.~Da Silva, P.~Richardson and C.~Wymant,
  Phys.\ Rev.\ D {\bf 86} (2012) 035023
  [arXiv:1203.3446 [hep-ph]].

\bibitem{Cerdeno:2009dv}
  D.~G.~Cerdeno and O.~Seto,
  JCAP {\bf 0908} (2009) 032
  [arXiv:0903.4677 [hep-ph]].


\bibitem{Cerdeno:2011qv}
  D.~G.~Cerdeno, J.~-H.~Huh, M.~Peiro and O.~Seto,
  JCAP {\bf 1111} (2011) 027
  [arXiv:1108.0978 [hep-ph]].


\bibitem{Huitu:2012rd}
  K.~Huitu, J.~Laamanen, L.~Leinonen, S.~K.~Rai and T.~Ruppell,
  JHEP {\bf 1211} (2012) 129
  [arXiv:1209.6302 [hep-ph]].

\bibitem{Borgani:1996ag}
  S.~Borgani, A.~Masiero and M.~Yamaguchi,
  Phys.\ Lett.\ B {\bf 386} (1996) 189
  [hep-ph/9605222].

\bibitem{Takayama:2000uz}
  F.~Takayama and M.~Yamaguchi,
  Phys.\ Lett.\ B {\bf 485} (2000) 388
  [hep-ph/0005214].

\bibitem{JeanLouis:2009du}
  C.~-C.~Jean-Louis and G.~Moreau,
  J.\ Phys.\ G {\bf 37} (2010) 105015
  [arXiv:0911.3640 [hep-ph]].

\bibitem{Luo:2010he}
  F.~Luo, K.~A.~Olive and M.~Peloso,
  JHEP {\bf 1010} (2010) 024
  [arXiv:1006.5570 [hep-ph]].

\bibitem{Craig:2012xp}
  N.~Craig, S.~Knapen, D.~Shih and Y.~Zhao,
  arXiv:1206.4086 [hep-ph].

\bibitem{Evans:2012bf}
  J.~A.~Evans and Y.~Kats,
  JHEP {\bf 1304} (2013) 028
  [arXiv:1209.0764 [hep-ph]].

\bibitem{Graham:2014vya}
  P.~W.~Graham, S.~Rajendran and P.~Saraswat,
  arXiv:1403.7197 [hep-ph].

\bibitem{CMS:2013qda}
  CMS Collaboration [CMS Collaboration],
  CMS-PAS-SUS-13-010.

\bibitem{ATLAS:2012dp}
  G.~Aad {\it et al.}  [ATLAS Collaboration],
  JHEP {\bf 1212} (2012) 086
  [arXiv:1210.4813 [hep-ex]].

\bibitem{Joshipura:1994wm}
  A.~S.~Joshipura and M.~Nowakowski,
  Phys.\ Rev.\ D {\bf 51} (1995) 5271
  [hep-ph/9403349].

\bibitem{romao1}
  J.~C.~Romao and J.~W.~F.~Valle,
  Phys.\ Lett.\ B {\bf 272} (1991) 436.

\bibitem{romao2}
  J.~C.~Romao, C.~A.~Santos and J.~W.~F.~Valle,
  Phys.\ Lett.\ B {\bf 288} (1992) 311.

\bibitem{Chaichian:1991zt}
  M.~Chaichian and A.~V.~Smilga,
  Phys.\ Rev.\ Lett.\  {\bf 68} (1992) 1455.

\bibitem{Chaichian:1992ra}
  M.~Chaichian and R.~Gonzalez Felipe,
  Phys.\ Rev.\ D {\bf 47} (1993) 4723.

\bibitem{Kuchimanchi:1993jg}
  R.~Kuchimanchi and R.~N.~Mohapatra,
  Phys.\ Rev.\ D {\bf 48} (1993) 4352
  [hep-ph/9306290].

\bibitem{Huitu:1994zm}
  K.~Huitu and J.~Maalampi,
  Phys.\ Lett.\ B {\bf 344} (1995) 217
  [hep-ph/9410342].

\bibitem{Barger:2008wn}
  V.~Barger, P.~Fileviez Perez and S.~Spinner,
  Phys.\ Rev.\ Lett.\  {\bf 102} (2009) 181802
  [arXiv:0812.3661 [hep-ph]].

\bibitem{LopezFogliani:2005yw}
  D.~E.~Lopez-Fogliani and C.~Munoz,
  Phys.\ Rev.\ Lett.\  {\bf 97} (2006) 041801
  [hep-ph/0508297].

\bibitem{Chikashige:1980ui}
  Y.~Chikashige, R.~N.~Mohapatra and R.~D.~Peccei,
  Phys.\ Lett.\ B {\bf 98} (1981) 265.

\bibitem{Gelmini:1980re}
  G.~B.~Gelmini and M.~Roncadelli,
  Phys.\ Lett.\ B {\bf 99} (1981) 411.

  
\bibitem{Abel:1995wk}
  S.~A.~Abel, S.~Sarkar and P.~L.~White,
  Nucl.\ Phys.\ B {\bf 454} (1995) 663
  [hep-ph/9506359].

\bibitem{Maniatis:2009re}
  M.~Maniatis,
  Int.\ J.\ Mod.\ Phys.\ A {\bf 25} (2010) 3505,
  arXiv:0906.0777 [hep-ph].
  
\bibitem{Abel:1996cr}
  S.~A.~Abel,
  Nucl.\ Phys.\ B {\bf 480} (1996) 55
  [hep-ph/9609323].
  
\bibitem{Panagiotakopoulos:1998yw}
  C.~Panagiotakopoulos and K.~Tamvakis,
  Phys.\ Lett.\ B {\bf 446} (1999) 224,
  hep-ph/9809475.

\bibitem{Panagiotakopoulos:1999ah}
  C.~Panagiotakopoulos and K.~Tamvakis,
  Phys.\ Lett.\ B {\bf 469} (1999) 145
  [hep-ph/9908351].


\bibitem{Frere:1983ag}
  J.~M.~Frere, D.~R.~T.~Jones and S.~Raby,
  Nucl.\ Phys.\ B {\bf 222} (1983) 11.
  
\bibitem{Drees:1988fc}
  M.~Drees,
  Int.\ J.\ Mod.\ Phys.\ A {\bf 4} (1989) 3635.
    
\bibitem{Gozdz:2008zz}
  M.~Gozdz and W.~A.~Kaminski,
  Phys.\ Rev.\ D {\bf 78} (2008) 075021
  [arXiv:1201.1241 [hep-ph]].

\bibitem{ATLAS:2012kr}
  G.~Aad {\it et al.}  [ATLAS Collaboration],
  JHEP {\bf 1212} (2012) 124
  [arXiv:1210.4457 [hep-ex]].

\bibitem{Marshall:2014kea}
  Z.~Marshall, B.~A.~Ovrut, A.~Purves and S.~Spinner,
  Phys.\ Lett.\ B {\bf 732} (2014) 325
  [arXiv:1401.7989 [hep-ph]].

\bibitem{Chatrchyan:2013xsw}
  S.~Chatrchyan {\it et al.}  (CMS Collaboration),
  arXiv:1306.6643 [hep-ex].

\bibitem{Wang:2013jya}
  W.~Wang, J.~M.~Yang and L.~L.~You,
  JHEP {\bf 1307} (2013) 158
  [arXiv:1303.6465 [hep-ph]].

\bibitem{Barate:2003sz}
  R.~Barate {\it et al.}  (LEP Working Group for Higgs boson searches and ALEPH and DELPHI and L3 and OPAL Collaborations),
  Phys.\ Lett.\ B {\bf 565} (2003) 61,
  hep-ex/0306033.

\bibitem{Badziak:2013bda}
  M.~Badziak, M.~Olechowski and S.~Pokorski,
  JHEP {\bf 1306} (2013) 043
  [arXiv:1304.5437 [hep-ph]].


\bibitem{Duggan:2013yna}
  D.~Duggan, J.~A.~Evans, J.~Hirschauer, K.~Kaadze, D.~Kolchmeyer, A.~Lath and M.~Walker,
  arXiv:1308.3903 [hep-ph].

\bibitem{Adam:2013mnn}
  J.~Adam {\it et al.}  (MEG Collaboration),
  arXiv:1303.0754 [hep-ex].

\bibitem{Bellgardt:1987du}
  U.~Bellgardt {\it et al.}  (SINDRUM Collaboration),
  Nucl.\ Phys.\ B {\bf 299} (1988) 1.

\bibitem{Ibarra:2011xn}
  A.~Ibarra, E.~Molinaro and S.~T.~Petcov,
  Phys.\ Rev.\ D {\bf 84} (2011) 013005
  [arXiv:1103.6217 [hep-ph]].

\bibitem{Dinh:2012bp}
  D.~N.~Dinh, A.~Ibarra, E.~Molinaro and S.~T.~Petcov,
  JHEP {\bf 1208} (2012) 125
   [Erratum-ibid.\  {\bf 1309} (2013) 023]
  [arXiv:1205.4671 [hep-ph]].

\bibitem{Mu3e}
Mu3e Collaboration, Andre Sch\"oning on behalf of the Mu3e Collaboration, {\it A New
Search for the Decay $\mu\rightarrow e^+e^+e^-$ }, contribution submitted to
European strategy in 2012, 
https://indico.cern.ch/abstractDisplay.py/getAttachedFile?abstractId=102\&resId=0\&confId=175067.

\bibitem{Blondel:2013ia}
  A.~Blondel, A.~Bravar, M.~Pohl, S.~Bachmann, N.~Berger, M.~Kiehn, A.~Schoning and D.~Wiedner {\it et al.},
  arXiv:1301.6113 [physics.ins-det].
  
\bibitem{HFG}
Heavy Flavor Averaging Group,
http://www.slac.stanford.edu/xorg/hfag/rare/2013/radll/index.html.



\bibitem{Deschamps:2009rh}
  O.~Deschamps, S.~Descotes-Genon, S.~Monteil, V.~Niess, S.~T'Jampens and V.~Tisserand,
  Phys.\ Rev.\ D {\bf 82} (2010) 073012
  [arXiv:0907.5135 [hep-ph]].
  


\bibitem{CMSandLHCbCollaborations:2013pla}
  CMS and LHCb Collaborations [CMS and LHCb Collaboration],
  CMS-PAS-BPH-13-007.

\bibitem{Buras:2013uqa}
  A.~J.~Buras, R.~Fleischer, J.~Girrbach and R.~Knegjens,
  JHEP {\bf 1307} (2013) 77
  [arXiv:1303.3820 [hep-ph]].
  
\bibitem{Li:2013fa}
  C.~Li, C.~-D.~Lu and X.~-D.~Gao,
  Commun.\ Theor.\ Phys.\  {\bf 59} (2013) 711
  [arXiv:1301.3445 [hep-ph]].
  
\bibitem{Dreiner:2013jta}
  H.~K.~Dreiner, K.~Nickel and F.~Staub,
  Phys.\ Rev.\ D {\bf 88} (2013) 115001
  [arXiv:1309.1735 [hep-ph]].
 
\bibitem{Babu:1999hn}
  K.~S.~Babu and C.~F.~Kolda,
  Phys.\ Rev.\ Lett.\  {\bf 84} (2000) 228
  [hep-ph/9909476].

\bibitem{Boos:2002ze}
  E.~Boos, A.~Djouadi, M.~Muhlleitner and A.~Vologdin,
  Phys.\ Rev.\ D {\bf 66} (2002) 055004
  [hep-ph/0205160].

\bibitem{Hall:1993gn}
  L.~J.~Hall, R.~Rattazzi and U.~Sarid,
  Phys.\ Rev.\ D {\bf 50} (1994) 7048
  [hep-ph/9306309, hep-ph/9306309].

\bibitem{Hempfling:1993kv}
  R.~Hempf\mbox{}ling,
  Phys.\ Rev.\ D {\bf 49} (1994) 6168.

\bibitem{Carena:1994bv}
  M.~S.~Carena, M.~Olechowski, S.~Pokorski and C.~E.~M.~Wagner,
  Nucl.\ Phys.\ B {\bf 426} (1994) 269
  [hep-ph/9402253].

\bibitem{Pierce:1996zz}
  D.~M.~Pierce, J.~A.~Bagger, K.~T.~Matchev and R.~-j.~Zhang,
  Nucl.\ Phys.\ B {\bf 491} (1997) 3
  [hep-ph/9606211].

\bibitem{Han:2013sga}
  C.~Han, X.~Ji, L.~Wu, P.~Wu and J.~M.~Yang,
  JHEP {\bf 1404} (2014) 003
  [arXiv:1307.3790 [hep-ph]].

\bibitem{Ellwanger:2011aa}
  U.~Ellwanger,
  JHEP {\bf 1203} (2012) 044,
  arXiv:1112.3548 [hep-ph].

\bibitem{Ferreira:2012my}
  P.~M.~Ferreira, R.~Santos, M.~Sher and J.~P.~Silva,
  Phys.\ Rev.\ D {\bf 85} (2012) 035020,
  arXiv:1201.0019 [hep-ph].

\bibitem{Baglio:2012et}
  J.~Baglio, A.~Djouadi and R.~M.~Godbole,
  Phys.\ Lett.\ B {\bf 716} (2012) 203,
  arXiv:1207.1451 [hep-ph].

\bibitem{Giudice:2012pf}
  G.~F.~Giudice, P.~Paradisi and A.~Strumia,
  arXiv:1207.6393 [hep-ph].

\bibitem{Gunion:2012gc}
  J.~F.~Gunion, Y.~Jiang and S.~Kraml,
  Phys.\ Rev.\ D {\bf 86} (2012) 071702
  [arXiv:1207.1545 [hep-ph]].

\bibitem{Agashe:2014kda}
  K.~A.~Olive {\it et al.}  [Particle Data Group Collaboration],
  Chin.\ Phys.\ C {\bf 38} (2014) 090001.


\bibitem{Djouadi:2013qya}
  A.~Djouadi and G.~Moreau,
  Eur.\ Phys.\ J.\ C {\bf 73} (2013) 2512
  [arXiv:1303.6591 [hep-ph]].

\bibitem{Belanger:2013xza}
  G.~Belanger, B.~Dumont, U.~Ellwanger, J.~F.~Gunion and S.~Kraml,
  Phys.\ Rev.\ D {\bf 88} (2013) 075008
  [arXiv:1306.2941 [hep-ph]].

\end{thebibliography}
\end{document}